\documentclass[twocolumn]{aastex631}

\usepackage[T1]{fontenc}
\usepackage{multirow}
\usepackage{hyperref}
\usepackage{graphicx}
\usepackage{amsmath}

\newcommand\ot{O$_{\mathrm{2}}$}
\newcommand\ee{$\eta_{\oplus}$}
\newcommand\re{$R_{\oplus}$}

\begin{document}

\title{Bioverse: GMT and ELT Direct Imaging and High-Resolution Spectroscopy Assessment -- Surveying Exo-Earth O$_{\mathrm{2}}$ and Testing the Habitable Zone Oxygen Hypothesis}

\author[0000-0003-3702-0382]{Kevin K.\ Hardegree-Ullman}
\affiliation{Steward Observatory, The University of Arizona, 933 N. Cherry Ave., Tucson, AZ 85721, USA}
\affiliation{Caltech/IPAC-NASA Exoplanet Science Institute, 1200 E. California Blvd., MC 100-22, Pasadena, CA 91125, USA}

\correspondingauthor{Kevin K.\ Hardegree-Ullman}
\email{kevinkhu@caltech.edu}

\author[0000-0003-3714-5855]{D\'{a}niel Apai}
\affiliation{Steward Observatory, The University of Arizona, 933 N. Cherry Ave., Tucson, AZ 85721, USA}
\affiliation{Lunar and Planetary Laboratory, The University of Arizona, 1629 E. University Blvd., Tucson, AZ 85721, USA}

\author[0000-0001-5130-9153]{Sebastiaan Y.\ Haffert}
\affiliation{Leiden Observatory, Leiden University, PO Box 9513, 2300, RA Leiden, The Netherlands}
\affiliation{Steward Observatory, The University of Arizona, 933 N. Cherry Ave., Tucson, AZ 85721, USA}

\author[0000-0001-8355-2107]{Martin Schlecker}
\affiliation{Steward Observatory, The University of Arizona, 933 N. Cherry Ave., Tucson, AZ 85721, USA}

\author[0000-0002-8425-6606]{Markus Kasper}
\affiliation{European Southern Observatory, Karl-Schwarzschild-Stra\ss{}e 2, 85748 Garching bei M{\"u}nchen, Germany}

\author[0000-0003-2769-0438]{Jens Kammerer}
\affiliation{European Southern Observatory, Karl-Schwarzschild-Stra\ss{}e 2, 85748 Garching bei M{\"u}nchen, Germany}

\author[0000-0002-4309-6343]{Kevin Wagner}
\affiliation{Steward Observatory, The University of Arizona, 933 N. Cherry Ave., Tucson, AZ 85721, USA}

\begin{abstract}

Biosignature detection in the atmospheres of Earth-like exoplanets is one of the most significant and ambitious goals for astronomy, astrobiology, and humanity. Molecular oxygen is among the strongest indicators of life on Earth, but it will be extremely difficult to detect via transmission spectroscopy. We used the \texttt{Bioverse} statistical framework to assess the ability to probe Earth-like O$_{\mathrm{2}}$ levels on hypothetical nearby habitable zone exo-Earth candidates (EECs) using direct imaging and high-resolution spectroscopy on the Giant Magellan Telescope (GMT) and the Extremely Large Telescope (ELT). Assuming continued improvement in instruments and data processing, our analysis highlights the best-case scenarios. Earth-like O$_{\mathrm{2}}$ levels could be probed on up to $\sim$7 and $\sim$19 EECs orbiting bright M~dwarfs within 20~pc in a hypothetical 10-year survey on the GMT and ELT, respectively. Four known super-Earth candidates, including Proxima Centauri~b, could be probed for O$_{\mathrm{2}}$ within about one week of observations on the ELT and a few months on the GMT. We also assessed the ability of the ELT to test the habitable zone oxygen hypothesis -- that habitable zone Earth-sized planets are more likely to have O$_{\mathrm{2}}$ -- within a 10-year survey using \texttt{Bioverse}. Testing this hypothesis requires either $\sim$1/2 of the EECs to have O$_{\mathrm{2}}$ or $\sim$1/3 if \ee\ is large. A northern hemisphere large-aperture telescope, such as the Thirty Meter Telescope (TMT), would expand the target star pool by about 25\%, reduce the time to probe biosignatures on individual targets, and provide an additional independent check on potential biosignature detections.

\end{abstract}


\section{Introduction} \label{sec:intro}

Methods for probing the atmospheres of exoplanets have matured to the point that we are within reach of biosignature detection on rocky worlds. Transmission spectroscopy has become a powerful tool to characterize the atmospheres of planets from the ground \citep[e.g.,][]{Snellen2008,Snellen2010,Stevenson2014}, and from space with HST \citep[e.g.,][]{Charbonneau2002,Sing2016,Zhang2018} and now JWST \citep[e.g.,][]{JWSTERS2023,Moran2023}. In the search for biosignatures via transmission spectroscopy, JWST might prove fruitful in the detection of species such as CH$_{\mathrm{4}}$, H$_{\mathrm{2}}$O, and CO$_{\mathrm{2}}$ \citep[e.g.,][]{Wunderlich2019}, however, it will not be able to detect \ot\ \citep{Lustig-Yaeger2019,Wunderlich2019,Pidhorodetska2020}, a major byproduct of the biological process of oxygenic photosynthesis on Earth. \citet{Snellen2013}, \citet{Rodler2014}, \citet{Serindag2019}, \citet{Lopez-Morales2019}, \citet{Currie2023}, and \citet{Hardegree-Ullman2023} studied the prospects of detecting \ot\ via transmission spectroscopy using upcoming 25--40-meter class telescopes such as the Giant Magellan Telescope \citep[GMT,][]{Johns2012}, Thirty Meter Telescope \citep[TMT,][]{Nelson2008,Sanders2013}, and European Southern Observatory Extremely Large Telescope \citep[ELT,][]{Gilmozzi2007}. \citet{Hardegree-Ullman2023} took into account practical ground-based observing constraints such as transit observability and relative system velocities to mitigate telluric blending. They concluded that even in a very optimistic observing scenario where signals from the GMT, TMT, and ELT and all observable transits could be combined, it will likely take decades to make an \ot\ detection on a single transiting habitable zone Earth-sized exoplanet.

This study explores whether direct imaging is more efficient for \ot\ surveys. Previous studies suggest it might be possible to detect biosignatures using reflected light spectroscopy on directly imaged planets \citep[e.g.,][]{Snellen2015,Lovis2017}. To date, high contrast imaging has been limited to young, self-luminous gas giants on wide orbits \citep[e.g.,][]{Chauvin2004,Marois2008,Lagrange2009,Macintosh2015,Keppler2018}. This is mainly due to telescope apertures limiting the inner working angle at which a planet is detectable (often similar to the Rayleigh criterion $\sim$1.22$\lambda/D$, where $\lambda$ is the observing wavelength and $D$ is the primary mirror diameter) and due to contrast limitations of current high-contrast imaging systems.

\subsection{Previous Direct Imaging Studies} \label{sec:previous}

There have been numerous studies investigating the potential for a next-generation space telescope \citep[e.g., LUVOIR, HabEx, LIFE,][]{TheLUVOIRTeam2019,Gaudi2020,Quanz2022} to probe biosignatures in reflected light with direct imaging and low-resolution spectroscopy \citep[e.g.,][]{Feng2018,Bixel2021,Damiano2022,Konrad2022,Robinson2023,Susemiehl2023}. Due to the recommendations brought forth in the Astro2020 Decadal Survey \citep{NationalAcademiesofSciences2021}, the Habitable Worlds Observatory (HWO) is being developed with the goal to identify and directly image at least 25 potentially habitable worlds using a telescope about the same size as JWST and optimistically launching in the 2040s. In the meantime, the GMT, TMT, and ELT are expected to be operational in the early 2030s. The significantly larger apertures of these telescopes will be able to probe smaller inner working angles and study the habitable zones of nearby M~dwarfs, which will be inaccessible to HWO.

Ground-based spectroscopy of biosignatures necessitates the use of high resolution spectrographs ($R\gtrsim50,000$) to allow separation of Earth's telluric lines from those of the exoplanet atmosphere, assuming sufficient line Doppler shifts to minimize line blending \citep[see, e.g.,][for examples of different resolutions and their effect on line blending]{Rodler2014,Lopez-Morales2019}.
\citet{Sparks2002} proposed combining a coronographic imager to a high spectral resolution integral-field spectrograph for detection and characterization of exoplanets. As \citet{Snellen2015} showed, the achievable contrast of a combined high-contrast imaging and high-resolution spectrograph system is the product of the individual system achievable contrasts. For example, if each individual system can reach a contrast of $\sim$10$^{-5}$, the combined system could reach contrasts of $\sim$10$^{-10}$ (see additional discussion in Section~\ref{sec:insmods}).

Notable observational developments have been made in this observing technique for measuring the optical albedo of hot Jupiters $\tau$~Bo{\"o}tis~b \citep[e.g.,][]{Charbonneau1999,Leigh2003,Rodler2014,Hoeijmakers2018} and 51~Pegasi~b \citep{Martins2015}. The first tentative detection came from \citet{Martins2015} who took 90 high-resolution spectra of 51~Pegasi~b with HARPS over the span of three months and used cross-correlation methods to measure a 3.7$\sigma$ upper limit to the planet-to-star contrast ratio of $6\times10^{-5}$, corresponding to a relatively high albedo of 0.5. This signal has been both confirmed \citep{Borra2018} and disputed \citep{DiMarcantonio2019,Scandariato2021,Spring2022}. \citet{Hoeijmakers2018} combined more than 2000 spectra collected over 15 years from four different facilities to measure a 3$\sigma$ upper limit to the planet-to-star contrast ratio of $1.5\times10^{-5}$ and a relatively low optical albedo of $\sim$0.12 for $\tau$~Bo{\"o}tis~b. These studies indicate the potential difficulty of using high-contrast imaging with high-resolution spectroscopy to detect planets in reflected light, but this could be due to the low albedo of hot Jupiters at optical wavelengths \citep{Brogi2021}. Measuring reflected-light signals from terrestrial planets may pose similar problems, but we optimistically continue to build and expand upon the techniques learned from reflected light studies of hot Jupiters.

\citet{Snellen2015} investigated using an optical high-resolution spectrograph (R=100,000) in combination with an extreme adaptive optics (AO) system on the ELT to observe a hypothetical 1.5~\re\ planet in the habitable zone of Proxima Centauri in reflected light \citep[prior to the discovery of Proxima Centauri~b by][]{Anglada-Escude2016}. They concluded that such a planet would be detectable in broadband (0.6--0.9~$\mu$m) reflected light in 10 hours of integration time at a S/N of $\sim$10 after cross-correlating the hypothetical observed spectrum with a model template spectrum, but they did not explore specific biosignature detection.

Since its discovery, most of the focus on high-contrast imaging and high-resolution reflected-light spectroscopy has been on Proxima Centauri~b. This is likely driven by the planet's location within the habitable zone, and a measured \textit{minimum mass} ($M \sin i$) likely between $1.0~M_{\oplus}$ \citep{Damasso2020} and $1.27~M_{\oplus}$ \citep{Anglada-Escude2016}. \citet{Bixel2017} placed probabilistic mass and radius constraints on Proxima Centauri~b at $M=1.63^{+1.66}_{-0.72}~M_{\oplus}$ and $R=1.07^{+0.38}_{-0.31}~R_{\oplus}$, which is consistent with a rocky composition, but there is still a $\sim$10\% chance the planet has a significant amount of ice or a volatile envelope. We do not yet have enough information to truly constrain the nature of Proxima Centauri~b, so assuming it is Earth-like or an Earth-analog is an optimistic assumption.

\citet{Lovis2017} proposed upgrades to SPHERE and ESPRESSO on the VLT in order to observe Proxima Centauri~b with high-resolution reflected-light spectra. They adopted a 3D Global Climate Model (GCM) atmosphere of Proxima Centauri~b from \citet{Turbet2016}, assumed observations at orbital quadrature, and a spectrograph resolution of 220,000. They concluded a 3.6$\sigma$ detection of \ot\ could be made in about 60 nights of telescope time over the course of three years.

\citet{Wang2017} simulated observations in reflected light with low to high-resolution spectra of Proxima Centauri~b (assuming it is an Earth-like planet) and a hypothetical Earth-like habitable zone planet orbiting an M~dwarf at 5~pc using a 30-meter class telescope at near-infrared bands. They used near-infrared (1--2.5~$\mu$m) R=1,000--100,000 Earth-like exoplanet spectra, considering atmospheric chemistry and using a radiative transfer model from \citet{Hu2012a,Hu2012b,Hu2013,Hu2014}. For a 100-hour integration time, their simulations yielded S/N$>$10 for H$_{\mathrm{2}}$O, \ot, CO$_{\mathrm{2}}$, and CH$_{\mathrm{4}}$, typically when R$>$10,000 and star-light suppression was $>10^{-7}$. These simulations simplistically addressed line blending, assuming the radial velocity of the exoplanet shifts its atmospheric signature by tens of km~s$^{-1}$. An updated version of this analysis was provided by \citet{Zhang2024}, where they simulated direct imaging of 10 nearby rocky planets at R=1000 with HARMONI-like and METIS-like instruments on the ELT in search of biosignatures. They indicated that CO$_{\mathrm{2}}$, CH$_{\mathrm{4}}$, and H$_{\mathrm{2}}$O should be detectable on GJ~887~b and Proxima Centauri~b at a S/N$>$5 with integration times on the order of $<$100 hours.

\citet{Hawker2019} focused on the detectability of Earth-like levels of \ot\ on Proxima Centauri~b with high-resolution spectroscopy and high contrast imaging on the ELT using HIRES and HARMONI-like instruments. They concluded a S/N=3--5 detection could be obtained in 30--70 hours of integration time. More recently, \citet{Vaughan2024} simulated observations of Proxima Centauri~b using HARMONI on the ELT at its highest resolving power of R=17,385. Their simulations indicated an atmosphere on Proxima Centauri~b (assuming it is Earth-like) could be characterized at S/N=5 in 20--30 hours of integration time. These observations would be particularly sensitive to CH$_{\mathrm{4}}$, but CO$_{\mathrm{2}}$ would be harder to detect.

While these previous studies have made significant contributions to assessing direct imaging capabilities for individual cases, they were intrinsically constrained in their scope for broader hypothesis testing. Our study seeks to build upon the above foundational efforts by employing the \texttt{Bioverse} framework, which allows us to go beyond previous studies by leveraging exoplanet demographics information, and performing realistic survey simulations and hypothesis testing.

\subsection{Bioverse} \label{sec:bioverse}

\texttt{Bioverse} was developed by \citet{Bixel2021} as a modular framework to generate and classify exoplanets based on exoplanet demographics, simulate exoplanet surveys for upcoming telescopes and missions, and compute the diagnostic power of future surveys in testing population-level hypotheses. \texttt{Bioverse} folds in state-of-the-art exoplanet occurrence rate calculations, incorporates a complete nearby star catalog out to $\sim$100~pc (Gaia red-band magnitude range $0.5<G_{RP}<19.9$), and accounts for realistic ground-based observing constraints, allowing exploration of the full potential of a broad statistical survey of nearby exoplanets.

The general workflow of \texttt{Bioverse} starts with generating synthetic populations of stars with planetary systems and to injecting a statistical trend into that population. Then, a simulated survey ``observes'' a sample of the synthetic planets and collects measurement data according to projected uncertainties and survey strategies. Next, the hypothesis testing module allows quantification of how well the injected trend is recovered, and how strongly parameters of the model shaping the trend can be constrained. This procedure is typically repeated in a Monte Carlo fashion, varying astrophysical unknowns or features of the survey, which allows for trade studies on different survey designs.

After introducing the framework in \citet{Bixel2021}, \citet{Hardegree-Ullman2023} added a stellar catalog based on Gaia Data Release~3, the capability to determine transit observability for ground-based large-aperture telescopes, and simulations for oxygen detection via transmission spectroscopy.
\citet{Schlecker2024} introduced selectable mass-radius relations, a mission simulator for ESA's PLATO mission~\citep{Rauer2016}, and a runaway greenhouse climate model.
Here, we built upon these previous studies to assess the direct imaging and high resolution spectroscopy potential for reflected light studies with the GMT and ELT, specifically focusing on the ability to probe Earth-like \ot\ levels. Leveraging all modules of \texttt{Bioverse}, we seek to test the ``habitable zone oxygen hypothesis,'' i.e., the hypothesis that Earth-sized planets within a star's habitable zone are more likely to have Earth-like O$_{\mathrm{2}}$ levels. This will inform us if a hypothetical 10-year survey of EECs with the ELT will be sufficient to test this hypothesis.

In Section~\ref{sec:obssim} we outline the direct imaging observing considerations and instrument models. We simulate a 10-year survey with the GMT and ELT to probe Earth-like levels of \ot\ on habitable zone Earths and super-Earths in Section~\ref{sec:survey}, and follow with testing the habitable zone oxygen hypothesis in Section~\ref{sec:hypothesis}. In Section~\ref{sec:discussion} we discuss the implications of our simulations, and we summarize and conclude our results in Section~\ref{sec:summary}.

\section{Observing Considerations}\label{sec:obssim}

Earth has evolved from a methane and carbon dioxide-rich atmosphere over 2.4 Gyr ago to its present-day oxygen-rich state due to oxygenic photosynthesis \citep[e.g.,][]{Lyons2021}. While it is likely that habitable zone Earth-sized exoplanets have atmospheres with diverse compositions and in different states of evolution, we adopt present-day Earth as a starting point to simplify simulations. The following simulations are focused on probing present-day Earth-like atmospheric conditions on rocky, habitable-zone exoplanets.

\subsection{Planet Generation and Observational Constraints}\label{sec:planets}

In order to provide a realistic survey simulation, we used \texttt{Bioverse} to generate a set of exoplanets orbiting bright stars (Gaia red-band $G_{RP}\leq10$) out to 20~pc from the Sun based on planet occurrence rates from \citet{Bergsten2022}, which was first introduced to \texttt{Bioverse} in \citet{Schlecker2024}. The underlying parametric model in \citet{Bergsten2022} is derived from planet radius, orbital period, and stellar host mass distributions in Kepler's FGK and early M~stars sample, considering planets with radii up to $3.5~R_{\oplus}$ and orbiting out to 100 days. The original SAG~13-based \citep{Kopparapu2018} planet occurrence rate module introduced by \citet{Bixel2021} was derived from a compilation of Kepler planet occurrence rates prior to 2017, and did not include a stellar mass dependence. Both of the planet occurrence rate modules are model-dependent for planets orbiting beyond 100 days. This does not affect our study much because an overwhelming majority of targets in our simulated surveys are M dwarfs (see Figure~\ref{fig:teff}) with habitable zones closer than 100 days. We note that the SAG~13 occurrence rates are more optimistic than the \citet{Bergsten2022} values by a factor of $\sim$2, but we opt to use the \citet{Bergsten2022} values in this study because it is not only more recent, but also from a uniform study rather than a compilation. 

From the set of generated exoplanets, we only considered planets that meet the criteria of an exo-Earth candidate (EEC): the planet is within the conservative habitable zone boundaries between runaway and maximum greenhouse as defined by \citet{Kopparapu2014}, and the planet radius falls between the range $0.8S^{0.25} < R_p < 1.4~R_{\oplus}$, where $S$ is stellar incident flux in units of present-day Earth insolation flux. The lower limit is the theoretical minimum planet size needed to retain an atmosphere \citep{Zahnle2017}, and was previously adopted for use in \texttt{Bioverse} by \citet{Bixel2021} and \citet{Hardegree-Ullman2023}.

In order to generate realistic surveys, for each EEC we simulated one Earth year of observations at one hour intervals and computed practical observing constraints based on the locations of the GMT and ELT. Simple constraints such as target observability at night and airmass between $z=1.0$ and 2.0 were modeled using \texttt{astroplan} \citep{astroplan2018}. When a planet was observable, we computed relative radial velocities and planet-to-star contrast ratios.

Since we have to observe \ot\ on exoplanets through the \ot-rich atmosphere of Earth, we must account for relative radial velocities ($\Delta$RV) in order to mitigate line blending effects. Relative radial velocities are given by:
\begin{equation}
    \Delta \mathrm{RV} = \mathrm{RV}_{\star}+\mathrm{RV}_{\mathrm{orb}}-\mathrm{RV}_{\mathrm{bary}},
\end{equation}
where $\mathrm{RV}_{\star}$ is the radial velocity of the host star, $\mathrm{RV}_{\mathrm{orb}}$ is the orbital velocity of the exoplanet with respect to its host star, and $\mathrm{RV}_{\mathrm{bary}}$ is the velocity of the observer on Earth with respect to the solar system barycenter. The host star radial velocity comes from the \texttt{Bioverse} stellar catalog~\citep{Hardegree-Ullman2023}, which is based on Gaia Data Release~3.
The uncertainties of these radial velocities are on the order of a few km~s$^{-1}$~\citep{GaiaCollaboration2021}.

Orbital velocity was computed from:
\begin{equation}
    \mathrm{RV}_{\mathrm{orb}} = \sqrt{GM_{\star} \left(\frac{2}{r}-\frac{1}{a} \right)},
\end{equation}
where $G$ is the gravitational constant, $M_{\star}$ is the mass of the host star, $r$ is the distance of the exoplanet to the host star at the time of observation (computed in \texttt{Bioverse} using Newton's method to numerically solve Kepler's equation), and $a$ is the semi-major axis of the exoplanet. Observer barycentric velocity at the time of observation was computed using the \texttt{baryCorr} function from the \texttt{PyAstronomy} package \citep{pya}. For an $R=100,000$ spectrograph observing the \ot\ A-band, severe line blending occurs when $|\Delta \mathrm{RV}|<13$ km s$^{-1}$ and $30\ \mathrm{km\ s}^{-1} < |\Delta \mathrm{RV}| < 55\ \mathrm{km\ s}^{-1}$ \citep{Lopez-Morales2019,Hardegree-Ullman2023}.

For direct imaging observations, the projected distance of the exoplanet from its host star ($\theta$ [arcsec] = $r$ [au]/$d$ [pc], where $d$ is the distance to the system) must be larger than the inner working angle of the observing instrument. The diffraction limit of a telescope is $1.22\lambda/D$. In practice, we set the observable inner working angle to be greater than $2\lambda/D$ \citep[e.g.,][]{Lovis2017,Walter2020}. At the wavelength of the \ot\ A-band (760 nm), $2\lambda/D$ is 13 mas, 11 mas, and 9 mas for the GMT, TMT, and ELT, respectively.

In order to observe the exoplanet, the contrast ratio between the planet and the star must be above the detection limit of the instrument. This contrast ratio is given by:
\begin{equation}
    C = \frac{F_p(\lambda,\alpha)}{F_{\star}(\lambda)} = A_g(\lambda) g(\alpha) \left(\frac{R_p}{a}\right)^2, \label{eq:fpfs}
\end{equation}
where $F_p(\lambda,\alpha)$ is the exoplanet flux at wavelength $\lambda$ and phase angle $\alpha$, $F_{\star}(\lambda)$ is the host star flux, $A_g$ is the geometric albedo, $g(\alpha)$ is the scattering phase function, $R_p$ is the exoplanet radius, and $a$ is the exoplanet semi-major axis. Our models assume a Lambertian phase function:
\begin{equation}
    g(\alpha) = \frac{\sin \alpha+(\pi-\alpha)\cos\alpha}{\pi},
\end{equation}
where the illumination phase angle $\alpha=\cos^{-1}(-\sin i \cos \phi)$, $i$ is the exoplanet inclination angle, and $\phi$ is the orbital phase. Geometric albedo and inclination angle were randomly generated from uniform distributions for each simulated planet within \texttt{Bioverse}, and orbital phase was computed at 1-hour intervals by summing the argument of periastron and the mean anomaly, assuming the longitude of the ascending node is 0 (in the direction toward the observer). The argument of periastron and mean anomaly are generated in \texttt{Bioverse} from uniform random distributions between 0 and 2$\pi$ and the mean anomaly is updated at 1-hour intervals, accounting for the simulated planet orbital period.

\subsection{Instrument Models}\label{sec:insmods}

In order to provide a realistic simulation of extreme-AO performance, we need models for planned instruments, which currently include GMagAO-X \citep{Males2022} for the GMT and the Planetary Camera and Spectrograph \citep[PCS,][]{Kasper2021} for the ELT. However, full end-to-end simulations of these high-order AO systems are very expensive in terms of computational time and complexity. Therefore, we used a semi-analytical approach to the performance simulation \citep{jolissaint2010synthetic}. The semi-analytical approach assumes that all temporal evolution of the turbulence is driven by Taylor's frozen flow. The frozen flow approximates the short time evolution by shifting the spatial structure of the turbulence with single-wind velocity. The time evolution of any parameter, $f$ for example, can then be replaced by a shifted version, $f(\vec{r}, t)=f(\vec{r} - \vec{v}t)$. This time evolution underpins the semi-analytical model and allows us to express an adaptive optics system as a series of spatial filters that act on the turbulence Power Spectral Density (PSD). The spatial frequency approach has been verified extensively in comparison with full end-to-end models \citep{males2021mysterious}. The AO system is controlled using integral control,
$\mathrm{DM}_{i+1} = \mathrm{DM}_{i} - g \varepsilon_i$. Here, $\varepsilon_i$ is the wavefront error at time step $i$ and $\mathrm{DM}_i$ is the DM shape at time step $i$. The feedback gain $g$ is a function of spatial frequency and was optimized following \citet{gendron1994astronomical, jolissaint2010synthetic}.

We simulated two different representative optimizations of the AO system. In the first, the AO system was allowed to run as fast as possible, which is 3~kHz for both GMagAO-X and PCS. The gain for each spatial frequency was optimized at this AO loop speed. In the second approach, we optimized the Strehl by changing both the AO loop speed and modal gains. These two approaches were used because optimizing for only Strehl does not necessarily lead to the best contrast. The contrast at small angular separations is driven by the servo lag error. This error term is a larger contributor on GMT/ELT-sized instruments than on current instruments \citep{males2018ground}. Therefore, it is possible to increase the contrast at smaller angular separations by running the AO system faster at the cost of lower Strehl. This leads to a better final signal-to-noise ratio. The two different approaches are called `Max Speed' and `Strehl Optimization' in this work. A more thorough assessment of which instrument and observing modes were selected for scientific analysis in Section~\ref{sec:survey} is given in Appendix~\ref{sec:app1}.

AO performance strongly depends on the atmospheric conditions assumed. For our simulations, we followed \cite{prieto2010giant} and \cite{thomas2010giant} for the GMT (see Table~\ref{tab:parameters}). The first quartile conditions have a Fried parameter of $r_0=0.22$~m and a $C_n^2$ integrated wind velocity of 9.4~m~s$^{-1}$. Here $C_n^2$ is the refractive index structure parameter. The second quartile conditions are $r_0=0.17$~m and a mean wind speed of 18.4 m~s$^{-1}$. For the ELT, we followed \cite{sarazin2013defining}: the first and second quartile conditions are $r_0=0.24$~m with a velocity of $10$ m~s$^{-1}$ and $r_0=0.15$~m with a velocity of 11 m~s$^{-1}$, respectively.

The design of GMagAO-X is currently further along than that of the other extreme-AO systems \citep{males2024high,haffert2024high}. Therefore, we used the current design parameters for GMagAO-X as reference. Specifically, we assumed 188 actuators across the pupil and a pyramid wavefront sensor. The design of PCS is currently still in its conceptual phase and no hard decisions have been made \citep{Kasper2021}. Therefore, we considered two different DM sizes for PCS: 200 actuators across the pupil based on the proposed density for EPICS \citep{kasper2010epics}, and 128 actuators across the pupil based on a more recent ELT extreme-AO design \citep{madec2022deformable}. 

\begin{figure*}[ht]
\centering
\includegraphics[width=\textwidth]{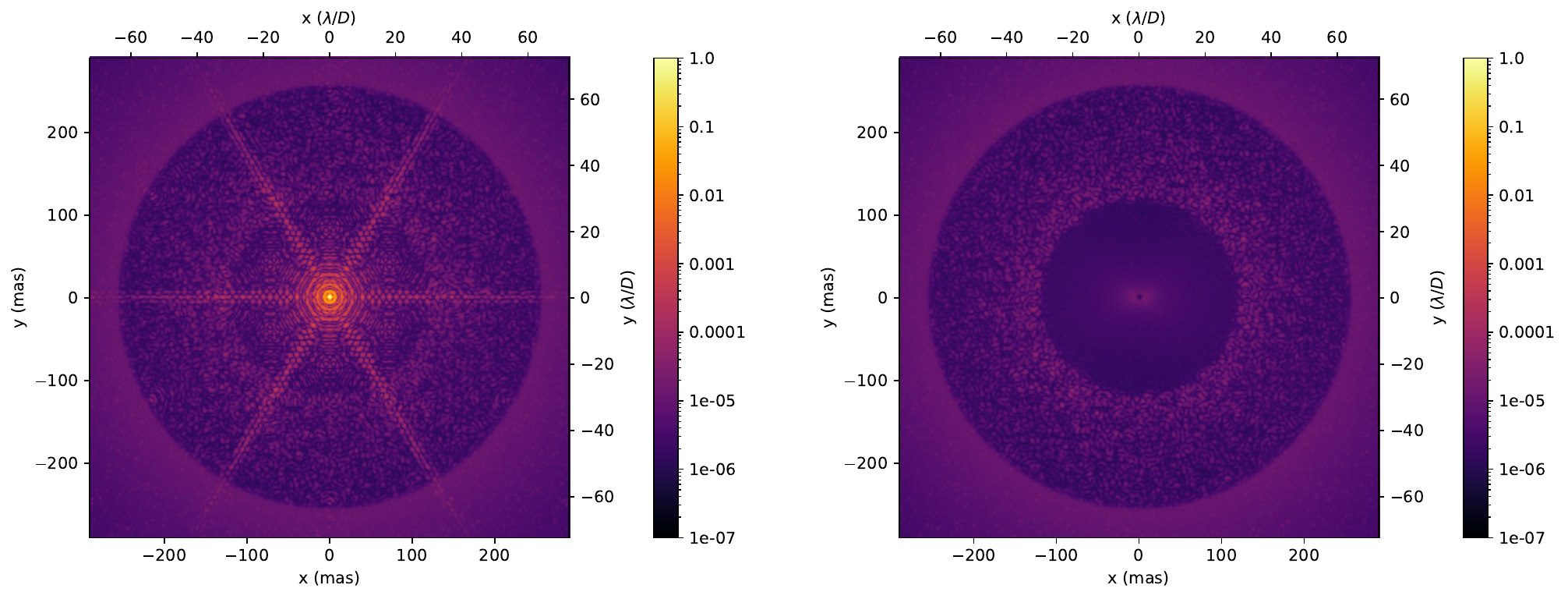}\\
\caption{Output of the semi-analytical AO simulations for ELT/PCS with the 128 actuators model. The left figure shows the simulated PSF and the right figure shows the coronagraphic stellar image. There are two visible circular regions. The largest corresponds to the control radius of the AO system and the smaller one corresponds to the control radius of the non-common path aberrations. The contrast within the inner radius is dominated by the wind-driven halo. Axes are shown in both mas and $\lambda/D$, assuming $\lambda=765$~nm and $D=39.3$~m.\label{fig:coronagraphic_image}}
\end{figure*}

The output of the semi-analytical model is a post-AO PSD. The PSD still has to be propagated through an actual coronagraphic system to measure the amount of residual scattered star light. The expected exposure times of our simulations are on the order of hours. This is much longer than the typical speckle lifetime of 1 to 10 ms \citep{males2021mysterious}. We can treat the different spatial frequencies as incoherent because of the many orders of magnitude differences in characteristic timescales. With that in mind, the propagation through a coronagraph can be viewed as an incoherent sum of plane waves \citep{sauvage2010analytical, herscovici2017analytic}. The theoretical perfect coronagraph removes only the piston mode from the incoming wavefront \citep{guyon2006theoretical, deshler2024achieving}. However, such coronagraphs are extremely sensitive to residual jitter and partially-resolved stars \citep{guyon2006theoretical}. Therefore, we used a 4th-order perfect coronagraph that also removes the tip/tilt modes from the wavefronts beside the piston mode. This increases the inner-working angle of the coronagraph and it increases the system robustness against stellar angular diameter and residual tip/tilt jitter\citep{guyon2006theoretical, belikov2021theoretical}. Given the size of the GMT/ELT, it will not be realistic to work with a 2nd-order coronagraph that only removes pistons. Finally, the effects of time varying quasi-static speckles are ignored in the simulations for two reasons. For high-resolution spectroscopy, the only thing that matters is the amount of starlight that is left in the final spectra. The cross-correlation function is differentiating between stellar residuals and planet light through spectral information only and the associated noise is photon noise of the time-integrated spectrum intensity \citep{Snellen2015}. The speckle lifetime affects the variance of this intensity but not the mean (shorter lifetime $\rightarrow$ smaller variance/mean). We assume that the quasi-static speckles can be controlled and removed to a level that is below the halo of residual atmospheric speckles using newly implemented focal plane wavefront control techniques \citep{potier2022increasing,haffert2023implicit, galicher2024increasing} and that means the photon noise from the atmospheric speckles will dominate the noise budget. Therefore, the total power of the low-spatial frequency content of the quasi-static speckles was set to 1~nm and the high-spatial frequency content was set to 25~nm and added to our simulations. This amount is similar to what has been achieved with MagAO-X \citep{VanGorkom2021}. We also assume that the efficiency of quasi-static speckle removal process will continue to improve over the next few years. An example output of the semi-analytical modeling is shown in Figure \ref{fig:coronagraphic_image}.

\begin{figure}[ht]
\centering
\includegraphics[width=\linewidth]{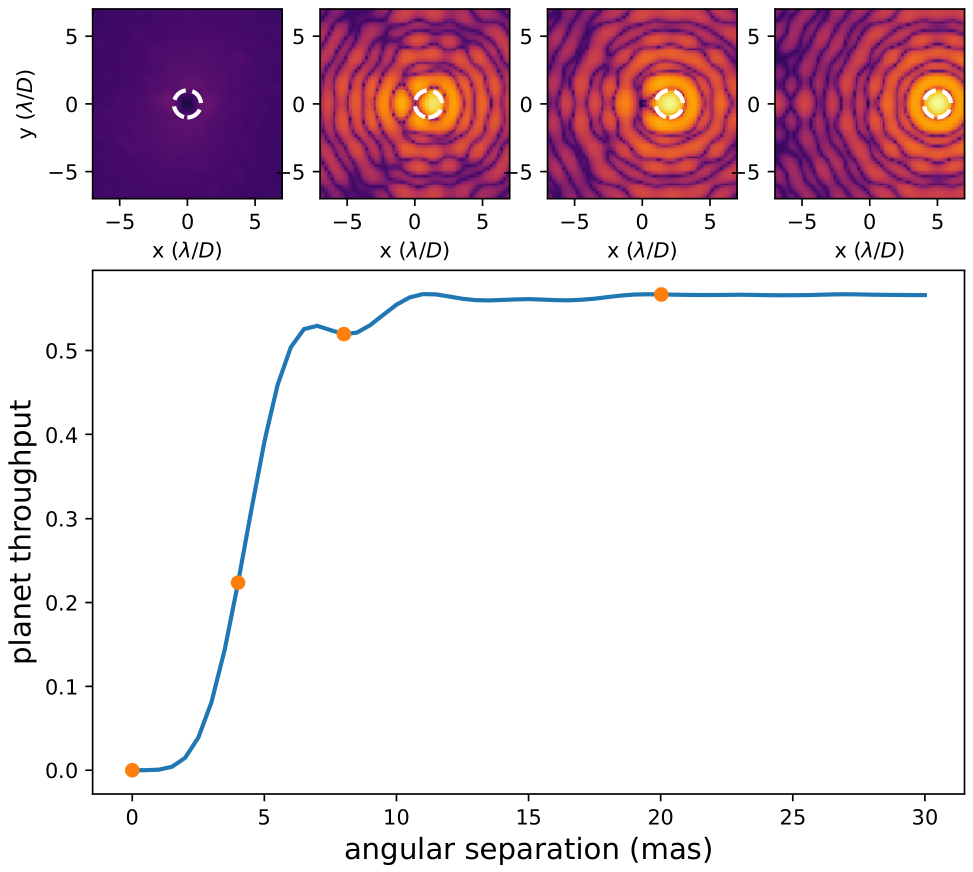}\\
\caption{Coronagraphic encircled energy throughput ($1.8 \lambda/D$ diameter) as a function of angular separation for an 8th magnitude star. A 4th-order perfect coronagraph with the ELT aperture was used in these simulations. The top row shows the PSFs corresponding to the four orange points on the throughput curve from 0 mas (left) to 20 mas (right) separation. The PSF is deformed close to the inner-working angle of the coronagraph. \label{fig:planet_throughput}}
\end{figure}

\begin{table*}[ht]
\begin{center}
\hskip-1.5cm\begin{tabular}{ll}
\hline \hline
ELT Parameters              &          \\ \hline \hline
Telescope area              &  978 m$^{2}$        \\
Number of actuators         &  128 / 200 \\
Number of wfs pixels        &  125,000  \\ \hline \hline
GMT Parameters              &   \\ \hline \hline
Telescope area              &  358 m$^{2}$        \\
Number of actuators         &  188     \\
Number of wfs pixels        &  111,000  \\ \hline \hline
Shared AO System Parameters              &          \\ \hline \hline
Zeropoint flux in I band    & 4520 photons s$^{-1}$ cm$^{-2}$ nm$^{-1}$ \\
Bandwidth I band    & 150 nm \\
WFS wavelength                 & 765 nm  \\ 
Loop speed        & 250 - 4000 Hz  \\
Intrinsic delay & 1.5 frames \\
Computational delay        & 250 $\mu$s \\
Throughput to WFS camera & 10 \% \\
WFS detector variance & 2.0 $e^{-}$\\
Photon noise sensitivity & $1/\sqrt{2}$ \\
Read noise sensitivity & $1/\sqrt{2}$ \\ 
Common path throughput to spectrograph  & 25 \%   \\ 
Coronagraph                & 4th order perfect coronagraph \citep{guyon2006theoretical}   \\ \hline \hline
Site Parameters            &          \\ \hline \hline
$r_0$                       & 10 - 20 cm    \\
$L_0$                       & 25 m     \\
$v$                         & 15 - 40 m s$^{-1}$    \\ \hline \hline
Spectrograph Parameters    &          \\ \hline \hline
read noise                 & $1 e^{-}$ pixel$^{-1}$   \\
dark current               & $10^{-4}$ $e^{-}$ s$^{-1}$ pixel$^{-1}$     \\
Spectrograph throughput      & 35 \%     \\
Detector quantum efficiency  & 86 \% \\
Science wavelength                 & 765 nm   \\
Spectral bandwidth         & 10 nm \\
Resolving power & 100,000 -- 500,000 \\
Spectral resolution        & 3 channels per resolving element \\
Pixels per channel         & 2 pixels \\ \hline
\end{tabular}
\end{center}
\caption{Parameters used to calculate S/N (Equation~\ref{eq:snr1}).}\label{tab:parameters}
\end{table*}

In order to calculate the signal-to-noise ratio of the high-contrast, high resolution spectroscopy observations of the exoplanet atmosphere \ot\ A-band, we follow the basic principle of \citet{Snellen2015} and \citet{Lovis2017}:

\begin{widetext}
\begin{eqnarray}\label{eq:snr1}
    S/N = \frac{\eta \Delta t T_p T_I T_{AO} T_{QE} C F_\star}{\sqrt{\Delta t T_\star T_I T_{AO} T_{QE} F_\star + \Delta t\sigma^2_{\mathrm{sky}} + N_{\mathrm{exp}} N_{\mathrm{pixels}}\sigma^2_{\mathrm{read}}+N_{\mathrm{pixels}}\Delta t\sigma^2_\mathrm{dark}}},
\end{eqnarray}
\end{widetext}
. The main difference between this equation and previous equations \citep{Snellen2015, Lovis2017} is the parametrization of the planet signal after post-processing. Earlier work used the combination of the average depth of the spectral line $\delta$ and the number of spectral lines $N_{\mathrm{lines}}$ to estimate the planet signal. However, this does not capture how post-processing affects the line-flux, which is the thing that we are sensitive to. High-resolution spectroscopy in direct imaging uses spectral high-pass filters to filter away stellar speckles. High-spectral resolution features are then mostly left alone. In our new approach, we measure the power of the planet signal before and after the spectral high-pass filtering and then use the ratio to estimate how much of the signal is retained. This ratio is the post-processing efficiency parameter $\eta$ \citep{Fowler2023}. At high-spectral resolution ($R>100.000$), the planetary oxygen line signal is almost completely retained. This leads to a high post-processing efficiency of 95\%. A caveat for this number is that it depends on the spectral bandwidth of the measurement. The signal that we measure is completely contained by oxygen spectral lines. Over a wide spectral bandwidth the efficiency drops because we remove all broadband continuum features (due to the spectral high-pass filter). Therefore, 95\% efficiency is achieved only in a 10 nm bandwidth centered around the oxygen A-band. The other parameters in the equation are: $\Delta t$ is the total exposure time, $T_p$ is the amount of light that is injected into a 1.8$\lambda / D$ sized light bucket (e.g., an optical fiber or IFU spaxel) after AO correction and through a coronagraph. Figure \ref{fig:planet_throughput} shows how this throughput varies with off-axis angle for an 8th magnitude primary star, and Table~\ref{tab:GMTthroughput-sm} shows how throughput varies for GMagAO-X for different magnitude primary stars. There are several places where flux is lost; the throughput losses in the spectrograph itself, $T_I$, the throughput through the common path optics of the AO system, $T_{AO}$, and the quantum efficiency of the detector $T_{QE}$. The planet-to-star contrast is $C$ (Equation~\ref{eq:fpfs}),  $F_\star$ is host star flux across the \ot\ A-band (10~nm bandwidth) at the top of Earth's atmosphere, $T_\star$ is stellar throughput measured at the location of the planet in the 1.8$\lambda/D$ sized aperture, and $\sigma^2_\mathrm{sky}$, $\sigma^2_\mathrm{read}$, and $\sigma^2_\mathrm{dark}$ are the sky background noise, detector read noise, and dark current noise, per pixel respectively. The spectra are sampled by $N_{\textrm{pixels}}$ per spectral channel and the observation is broken up into $N_{\textrm{exp}}$ exposures per hour of observing time. Our approach to the analytical modeling of high-resolution spectroscopy almost exactly follows the work in \citet{landman2023trade, Fowler2023, bidot2024exoplanet}. \citet{bidot2024exoplanet} also find that semi-analytical modeling of the cross-correlation post-processing results in the exact same performance as full end-to-end modeling. We used the designs of the RHEA and RISTRETTO spectrographs \citep{rains2016precision, bugatti2024ristretto} as a guide for our chosen spectrograph performance requirements (see Table~\ref{tab:parameters}).

At high resolution ($R\geq100,000$), we can mitigate Earth's telluric signal by avoiding observing the exoplanet at relative radial velocities at which line blending is severe (Section~\ref{sec:planets}). Disentangling the star and exoplanet signal happens in post-processing by cross-correlating the reflected-light spectrum with the stellar spectrum. Cross-correlation post-processing efficiency ($\eta$) for the \ot\ A-band converges to 95\% for $R\geq100,000$ \citep{Fowler2023} if the planet is sufficiently Doppler-shifted \citep{Hardegree-Ullman2023}.
Rearranging Equation~\ref{eq:snr1}, we computed the achievable contrast $C$ at S/N=5 for 1-hour total integration times (assuming four 15-minute exposures) for different host star apparent magnitudes and angular separations for the GMT (Table~\ref{tab:GMTcontrast-sm}) and ELT (Table~\ref{tab:ELTcontrast-sm}). The brightest stars might require shorter exposures, but the dominant detector noise will be dark current, which scales with exposure time, rather than read noise, so the effect is insignificant.

\begin{figure*}[ht]
\centering
\includegraphics[width=\textwidth]{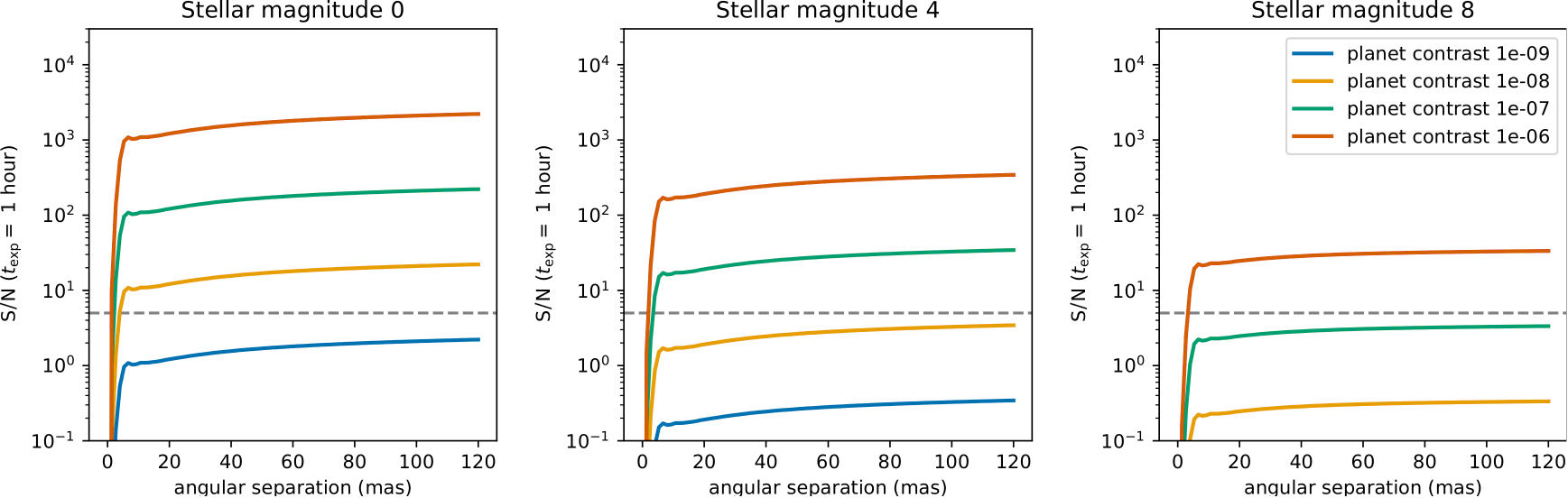}\\
\caption{The S/N after a 1 hour integration for different planet contrasts and stellar magnitudes. The S/N=5 detection limit is shown by the dashed grey line. \label{fig:snr_limits}}
\end{figure*}

There are several caveats to these simulations. First, there is no variability of the atmospheric conditions during a single observation. Variability is expected to lower the effective performance. However, we simulated the end-to-end yield in Q1 and Q2 conditions and these are used as upper and lower bounds on the performance. Later results will also show that the unknowns in the astrophysical statistics are dominating the yield estimates. Secondly, we assume that we reach the photon or detector noise dominated regime. This means that we assume little to no covariance between the spectral channels. This might seem optimistic, but \citet{Hoeijmakers2018} reached a post-processing contrast of $\sim$$10^{-5}$ by combining a significant amount of data from many different seeing-limited instruments, while the proposed survey will need to reach $10^{-3}$ to $10^{-4}$. Reaching the noise limits does require correct O$_2$ models for the cross-correlation because model imperfections will lead to lower efficiency \citep{Vaughan2024}. However, the O$_2$ transmission spectrum of the HITRAN model is generally very accurate because it predicts the atmospheric tellurics to a high precision \citep{GORDON2022}.

\subsection{Time Needed to Probe \texorpdfstring{O$_{\mathrm{2}}$}{O2}}\label{sec:time}
With our new instrument model, we now have to determine how long it will take to probe \ot. For each simulated exoplanet, we first computed the total number of hours the planet would be observable from the GMT and ELT locations for one year of observations ($N_{\text{hr-obs}}$) accounting for the aforementioned ground-based observing limitations and exoplanet orbital properties. Since the modeled planets change illumination and scattering phase throughout their orbits, we computed the contrast ratio from Equation~\ref{eq:fpfs} at each observable hour and used the resulting median contrast ratio ($\widetilde{F_p/F_{\star}}$) for our calculations. Tables~\ref{tab:GMTcontrast-sm} and \ref{tab:ELTcontrast-sm} show the expected achievable post-processing contrasts from the instrument models in Section~\ref{sec:insmods} for the GMT and ELT, respectively, at S/N = 5 for a one hour integration of the \ot\ A-band at $R=100,000$ for different host magnitudes and angular separations.

It would be ideal to optimize observations based on maximizing Equation~\ref{eq:fpfs}. However, this depends on several factors, including well constrained exoplanet orbits and phases and appropriate relative system velocities coinciding with Q1 or Q2 observing windows. It is impractical to assume observations will only happen when all these factors are most optimal, so we opted for a more conservative median observing situation. For each simulated exoplanet, we interpolated the host star $G_{RP}$ magnitude and median angular separation ($\tilde{\theta}$) to these tables to yield a 1-hour contrast ($C_{1\text{hr}}$). Total integration time to achieve a S/N = 5 detection of Earth-like levels of \ot\ on the simulated planet ($t_{\text{hr-}5\sigma}$) is given by:
\begin{equation}
    t_{\text{hr-}5\sigma} = \left(\frac{C_{1\text{hr}}}{\widetilde{F_p/F_{\star}}}\right)^2.
\end{equation}
To translate total integration time to years of observations, we need to account for both target observability ($N_{\text{hr-obs}}$) and acceptable site observing conditions ($Q$):
\begin{equation}
    t_{\text{yr-}5\sigma}  = Q\frac{t_{\text{hr-}5\sigma}}{N_{\text{hr-obs}}},
\end{equation}
where $Q=4$ for first quartile observing conditions (best observing site weather, ideal for extreme-AO, $\sim$25\% of telescope observing time, will take four times longer to achieve S/N) and $Q=2$ for second quartile observing conditions (best to average observing site weather, $\sim$50\% of telescope observing time, will take twice as long to achieve S/N). It is worth emphasizing that $N_{\text{hr-obs}}$ only counts times at which the relative system velocities yield unblended lines and the planet is observable from the observing site. We simulate first and second quartile observing conditions for each instrument mode, and assess the trade-offs of different observing scenarios in the Appendix.

\begin{table*}[ht]
\scriptsize
\begin{center}
\hskip-2.5cm\begin{tabular}{|c|cccccc|} \hline
        & \multicolumn{6}{c|}{GMT, Max Speed, Q1}  \\ \cline{2-7}
Mag|$\theta$ & 5 mas    & 10 mas   & 15 mas   & 30 mas   & 50 mas   & 120 mas \\ \hline
1 & 0.57 & 3.43 & 3.63 & 3.78 & 3.84 & 3.84 \\
3 & 0.57 & 3.40 & 3.60 & 3.74 & 3.80 & 3.81 \\
5 & 0.55 & 3.29 & 3.48 & 3.63 & 3.68 & 3.69 \\
7 & 0.48 & 2.90 & 3.07 & 3.19 & 3.24 & 3.25 \\
9 & 0.30 & 1.78 & 1.88 & 1.96 & 1.99 & 1.99 \\
10 & 0.16 & 0.94 & 1.00 & 1.04 & 1.06 & 1.06 \\ \hline

\end{tabular}
\end{center}
\caption{End-to-end GMagAO-X planet throughput in percentages (from 0\% to 100\%) for different host star magnitudes and planet-star angular separations ($\theta$). This table is for a Max Speed AO control system and Q1 observing conditions. For different AO control systems and observing conditions see Table~\ref{tab:GMTcontrast-full} in Appendix~\ref{sec:app1}.}\label{tab:GMTthroughput-sm}
\end{table*}

\begin{table*}[ht]
\scriptsize
\begin{center}
\hskip-2.5cm\begin{tabular}{|c|cccccc|} \hline
        & \multicolumn{6}{c|}{GMT, Max Speed, Q1}  \\ \cline{2-7}
Mag|$\theta$ & 5 mas    & 10 mas   & 15 mas   & 30 mas   & 50 mas   & 120 mas \\ \hline
1 & 3.06E-08 & 6.09E-09 & 5.85E-09 & 4.53E-09 & 3.51E-09 & 2.35E-09 \\
3 & 9.79E-08 & 1.97E-08 & 1.93E-08 & 1.61E-08 & 1.32E-08 & 9.50E-09 \\
5 & 4.01E-07 & 8.10E-08 & 8.00E-08 & 6.88E-08 & 5.83E-08 & 4.39E-08 \\
7 & 2.26E-06 & 4.57E-07 & 4.54E-07 & 4.00E-07 & 3.47E-07 & 2.71E-07 \\
9 & 2.12E-05 & 4.30E-06 & 4.30E-06 & 3.87E-06 & 3.43E-06 & 2.72E-06 \\
10 & 8.87E-05 & 1.80E-05 & 1.81E-05 & 1.65E-05 & 1.49E-05 & 1.15E-05 \\ \hline
\end{tabular}
\end{center}
\caption{GMT achievable $F_p/F_{\star}$ contrast at S/N=5 for 1 hour integration times using high contrast imaging with high-resolution spectroscopy for different host star magnitudes and planet-star angular separations ($\theta$). This table is for a Max Speed AO control system and Q1 observing conditions. For different AO control systems and observing conditions see Table~\ref{tab:GMTcontrast-full} in Appendix~\ref{sec:app1}.}\label{tab:GMTcontrast-sm}
\end{table*}

\begin{table*}[ht]
\begin{center}
\scriptsize
\hskip-2.5cm\begin{tabular}{|c|cccccc|} \hline
& \multicolumn{6}{c|}{ELT, Max Speed, 128 actuators, Q1}                \\ \cline{2-7}
Mag|$\theta$ & 5 mas    & 10 mas   & 15 mas   & 30 mas   & 50 mas   & 120 mas  \\ \hline
1          & 2.55E-09 & 2.08E-09 & 1.99E-09 & 1.55E-09 & 1.25E-09 & 9.36E-10 \\
3          & 7.12E-09 & 5.84E-09 & 5.64E-09 & 4.64E-09 & 3.90E-09 & 2.95E-09 \\
5          & 2.48E-08 & 2.04E-08 & 1.99E-08 & 1.68E-08 & 1.43E-08 & 1.09E-08 \\
7          & 1.03E-07 & 8.49E-08 & 8.30E-08 & 7.14E-08 & 6.17E-08 & 4.78E-08 \\
9          & 5.72E-07 & 4.72E-07 & 4.64E-07 & 4.06E-07 & 3.57E-07 & 2.82E-07 \\
10         & 1.62E-06 & 1.34E-06 & 1.32E-06 & 1.17E-06 & 1.03E-06 & 8.24E-07 \\ \hline
\end{tabular}
\caption{ELT achievable $F_p/F_{\star}$ contrast at S/N=5 for 1 hour integration times using high contrast imaging with high-resolution spectroscopy for different host star magnitudes and planet-star angular separations ($\theta$). This table is for a Max Speed AO control system, 128 deformable mirror actuators, and Q1 observing conditions. For different AO control systems, actuators, and observing conditions see Table~\ref{tab:ELTcontrast-full} in Appendix~\ref{sec:app1}.}\label{tab:ELTcontrast-sm}
\end{center}
\end{table*}

\section{Survey Simulation}\label{sec:survey}
\begin{figure}[ht]
\centering
\includegraphics[width=\linewidth]{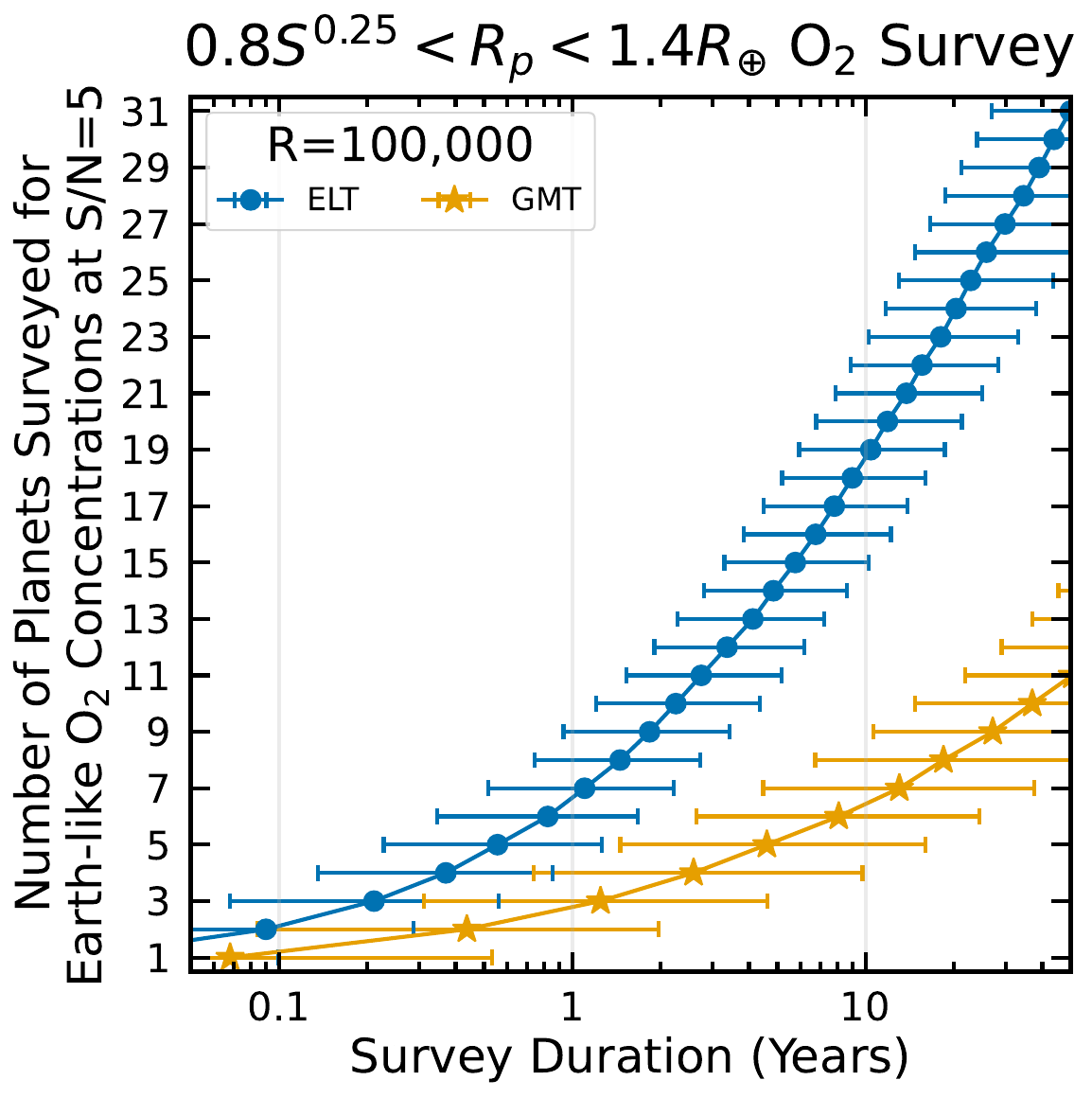}
\caption{Simulation results for a survey of Earth-like concentrations of \ot\ at S/N=5 on exo-Earth candidates within 20~pc with the ELT (blue dots) and the GMT (orange stars). A survey of the same duration on both telescopes would allow the ELT to probe about three times more planets for \ot\ than the GMT. The large error bars are driven by the wide range of possible orbital parameters for our simulated exoplanets. \label{fig:sims}}
\end{figure}

In order to realistically simulate the execution of an observing survey of nearby EECs, we followed the survey simulation methodology of \citet{Hardegree-Ullman2023}. First, we generated 1000 simulated local neighborhood ($d<20$~pc) exoplanet populations in \texttt{Bioverse} based on exoplanet demographics of \citet{Bergsten2022} and computed the time it will take to probe Earth-like \ot\ levels for each simulated EEC. We then sorted each population of simulated exoplanets by the number of years it would take to probe \ot\ and calculated the median value of the first through nth planets. The results are shown in Figure~\ref{fig:sims} for optimized instrument and observing modes, which are Max Speed, Q1 for the GMT, and Max Speed, Q1, 128 actuators for the ELT. An assessment of different instrument and observing modes is given in Appendix~\ref{sec:app1}.

\begin{figure}[ht]
\centering
\includegraphics[width=\linewidth]{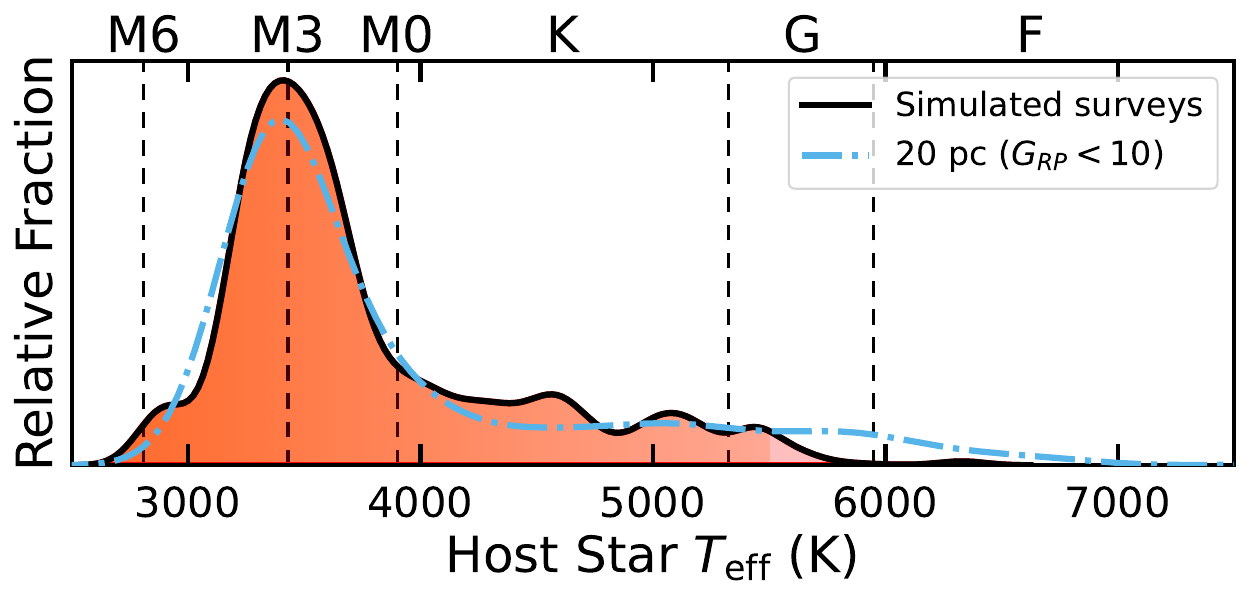}\\
\includegraphics[width=\linewidth]{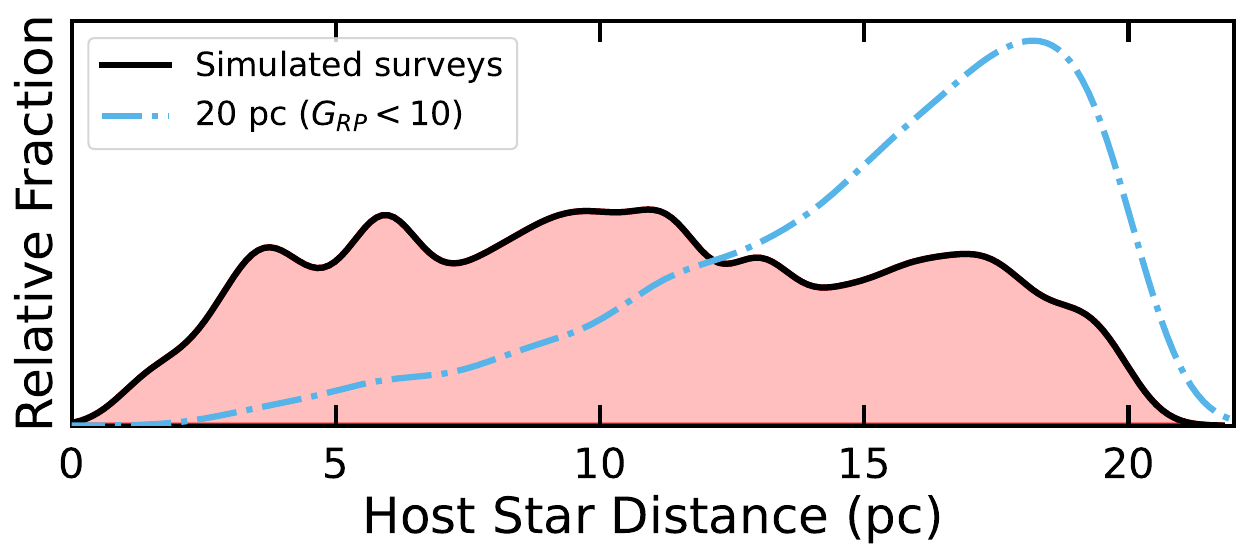}
\caption{Distributions of host star $T_{\mathrm{eff}}$ (top) and distance (bottom) across all EEC survey simulations for EECs that could be probed for Earth levels of \ot\ within a 10-year survey. 85\% of the detectable simulated planets orbit M dwarfs, with a peak near spectral type M3. About 3/4 of the detectable simulated planets are within 13~pc of the Sun. The dashed-dotted blue lines show the distribution of all bright ($G_{RP}<10$) stars within 20~pc, which highlights that the GMT and ELT will be insensitive to \ot\ in the atmospheres of EECs orbiting stars above $\sim$5000~K, and is most sensitive to stars within about 13~pc. \label{fig:teff}}
\end{figure}

Figure~\ref{fig:sims} shows that the ELT would be able to survey $\sim$3$\times$ more planets than the GMT for a survey of similar length, approximately commensurate with the $\sim$3$\times$ more collecting area. For example, in a 10-year survey from start to finish, if all EECs have Earth-like levels of \ot, the GMT could probe up to $\sim$7 planets, and the ELT could probe up to $\sim$19. Note, that this simulation did not consider that these planets would likely have overlapping observing windows, so the number of planets surveyed for \ot\ is an upper limit. Further observing optimization (e.g., observing exoplanets near maximum phase, or combining measurements from multiple telescopes) should be possible, but was not explored in this study. It is also important to highlight that over 85\% of the simulated planets which could be probed within a 10-year survey across all survey simulations orbit M dwarfs, and about 75\% of these simulated planets are within 13~pc of the Sun (Figure~\ref{fig:teff}).

\vspace{16pt}
\subsection{Super-Earths}\label{sec:se}
\begin{figure}[ht]
\centering
\includegraphics[width=\linewidth]{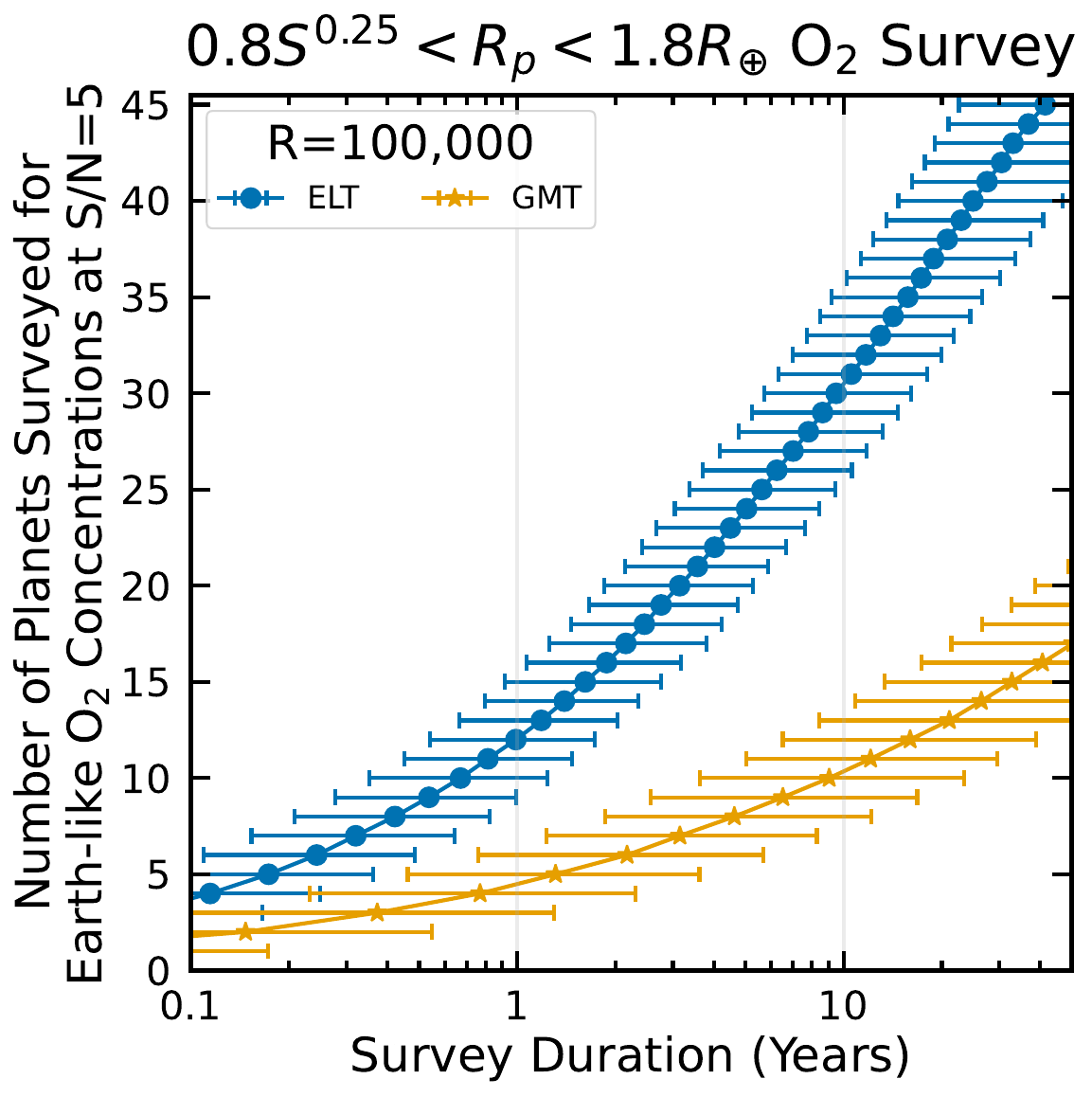}
\caption{Simulation results for a survey of Earth-like concentrations of \ot\ at S/N=5 on habitable zone Earth and super-Earth-sized candidates ($0.8S^{0.25} < R_p < 1.8~R_{\oplus}$) within 20~pc with the ELT (blue dots) and the GMT (orange stars). Similar to the exo-Earth simulation in Figure~\ref{fig:sims}, the ELT could probe about three times more planets than the GMT. Overall, a super-Earth survey could probe about 1.5 times more planets than a survey of planets up to 1.4~\re. \label{fig:sims-se}}
\end{figure}

Next, we carried out a simulation to explore the observability of \ot\ in all super-Earth exoplanets. The previous survey simulation focused on planets nearly the same size as Earth ($0.8S^{0.25} < R_p < 1.4~R_{\oplus}$). While not all exoplanets smaller than $1.4~R_{\oplus}$ will necessarily be Earth-like, a larger population including all super-Earth exoplanets would allow us to observationally constrain the limits of Earth-like planet atmospheres. In the following simulation, we assessed the ability to probe Earth-like \ot\ levels on Earth to super-Earth sized planets ($0.8S^{0.25} < R_p < 1.8~R_{\oplus}$), where the upper limit is guided by the exoplanet radius valley \citep[][]{Fulton2017,Hardegree-Ullman2020}. Results from the super-Earth survey simulation are given in Figure~\ref{fig:sims-se}. We found that about 1.5$\times$ more planets could be surveyed if the survey is broadened from Earth-sized to all Earth and super-Earth-sized planets.

\subsection{Known Habitable Zone Exoplanets}\label{sec:known}

\begin{figure}[ht]
\centering
\includegraphics[width=\linewidth]{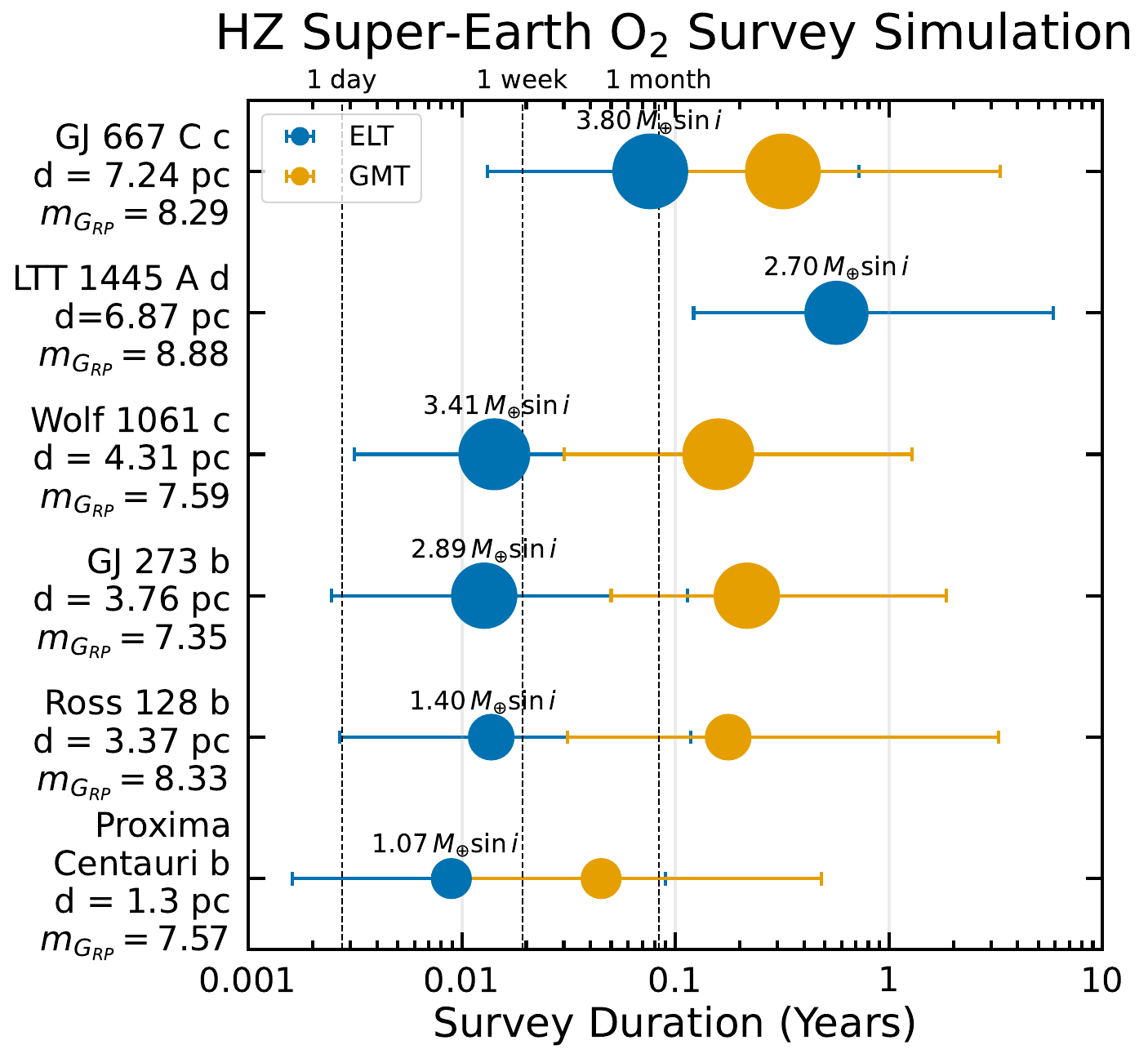} \\
\caption{Simulation results for a survey of Earth-like concentrations of \ot\ at S/N=5 for five known, and one candidate (LTT~1445~A~d) habitable zone exoplanets within 20~pc with the ELT (blue) and the GMT (orange). The ELT and GMT should both be able to probe for \ot\ on Proxima Centauri~b in less than a month. LTT~1445~A~d is the only planet in the simulation for which Earth-like levels of \ot\ could not be probed with the GMT, but it could be probed with the ELT. More precise planet and orbital parameters would minimize the uncertainties on these observing simulations. The minimum mass ($M_{\oplus}\sin i$) of each planet is given above each relatively-sized point \citep{Faria2022,Bonfils2018,Astudillo-Defru2017,Lavie2023,Anglada-Escude2013}. \label{fig:sims-real}}
\end{figure}

Within 20~pc there are five confirmed habitable zone exoplanets which would be observable from either the GMT or ELT. These exoplanets orbit stars brighter than ${G_{RP}}=10$ and have minimum masses ($M\sin i$) that make them super-Earth candidates. These are Proxima~Centauri~b \citep{Anglada-Escude2016}, Ross~128~b \citep{Bonfils2018}, GJ~273~b \citep{Astudillo-Defru2017}, Wolf~1061~c \citep{Wright2016}, and GJ~667~C~c \citep{Anglada-Escude2012}. LTT~1445~A~d is another nearby candidate super-Earth habitable zone exoplanet with a relatively bright host star \citep{Lavie2023}. It is worth noting that all these planets were discovered orbiting M~dwarfs using the radial velocity technique, and it is possible that their mass is significantly higher than the minimum mass provided by the RV method \citep[see e.g.,][]{Bixel2017}. We ran a survey simulation on these planets, assuming uniform priors on all unknown or loosely constrained parameters such as albedo, inclination angle, argument of periastron, and longitude of the ascending node. Results from this simulation are shown in Figure~\ref{fig:sims-real}. If Earth-like levels of \ot\ exist on these planets, it could be probed within days to a few weeks with the ELT, and within a couple of months with the GMT. For LTT~1445~A~d, it will take several months, and can only be completed on the ELT within a 10 year survey. Given the fact that the existence of these planets is already established, they will likely be high-priority targets for initial exoplanet surveys with the GMT and ELT.

\section{Hypothesis Testing}\label{sec:hypothesis}

Since our survey simulations revealed the possibility of detecting \ot\ on several exoplanets in a 10-year survey, particularly with the ELT, in order to assess the scientific value of such surveys, we now turn to testing related population-level hypotheses. We refer the reader to Section~5 of \citet{Bixel2021} for a full description of Bayesian hypothesis testing with \texttt{Bioverse}. In general, we want to assess the statistical power of a survey to test a hypothesis by running survey simulations several times in a Monte Carlo fashion under the same assumptions (but with randomly drawn parameters) and determine the fraction of simulations that validate or reject the hypothesis.
Here, we propose and aim to test the ``habitable zone oxygen hypothesis'': We hypothesize that habitable zone Earth-sized exoplanets will be more likely to have Earth-like levels of \ot{} than planets outside the habitable zone. 

We must first define the statistical effect to inject into our synthetic planet population. 
In our model, a planet can have or not have Earth-like levels of \ot.
It can further be an EEC if its orbit fulfills $a_{\mathrm{inner}} < a < a_{\mathrm{outer}}$ and its radius fulfills $0.8S^{0.25} < R_p < 1.4~R_{\oplus}$.
The planet is considered a non-EEC otherwise, but its atmosphere may still contain Earth-like levels of oxygen.
This population of oxygen-bearing non-EECs can be considered false positives, and they present a source of noise in the hypothesis testing.

We can now describe the likelihood of a planet having \ot\ by 
\begin{equation}
    f^{\mathrm{O}_\mathrm{2}} = \begin{cases}
        f^{\mathrm{O}_\mathrm{2}}_{\mathrm{EEC}} & \mathrm{if\ } a_{\mathrm{inner}} < a < a_{\mathrm{outer}} \\
        & \mathrm{and\ } 0.8S^{0.25} < R_p < 1.4~R_{\oplus} \\
        f^{\mathrm{O}_\mathrm{2}}_{\text{non-EEC}}  & \mathrm{if\ } a < a_{\mathrm{inner}} \mathrm{\ or\ } a > a_{\mathrm{outer}} \\
        & \mathrm{and\ }  R_p > 0.8S^{0.25} \\
        0 & \mathrm{if\ } R_p < 0.8S^{0.25},
    \end{cases}
\end{equation}
where $f^{\mathrm{O}_\mathrm{2}}_{\mathrm{EEC}}$ is the fraction of EECs with Earth-like \ot\ levels, and $f^{\mathrm{O}_\mathrm{2}}_{\text{non-EEC}}$ is the fraction of non-habitable zone exoplanets with \ot.

Next, we adopt the functional form of the habitable zone oxygen hypothesis, $H_{\mathrm{HZ}}^{\mathrm{O}_\mathrm{2}}$, as
\begin{equation}
    H_{\mathrm{HZ}}^{\mathrm{O}_\mathrm{2}}(a_{\mathrm{eff}}) = \begin{cases}
        f_{\mathrm{HZ}}^{\mathrm{O}_\mathrm{2}} \ \ \ \mathrm{if\ } a_{\mathrm{inner}} < a_{\mathrm{eff}} < a_{\mathrm{inner}}+\Delta a \\
        f_{\mathrm{HZ}}^{\mathrm{O}_\mathrm{2}}(f_{\text{non-HZ}}^{\mathrm{O}_\mathrm{2}}/f_{\mathrm{HZ}}^{\mathrm{O}_\mathrm{2}}) \ \ \mathrm{otherwise.}
    \end{cases}
\end{equation}
The model parameter vector $\theta$ consists of the position of the inner edge of the habitable zone $a_{\mathrm{inner}}$, the width of the habitable zone $\Delta a$, the fraction of habitable zone planets with \ot\ $f_{\mathrm{HZ}}^{\mathrm{O}_\mathrm{2}}$, and the ratio of non-habitable zone planets with \ot\ to habitable zone planets with \ot\ $f_{\text{non-HZ}}^{\mathrm{O}_\mathrm{2}}/f_{\mathrm{HZ}}^{\mathrm{O}_\mathrm{2}}$.

\begin{figure*}[ht!]
\centering
\includegraphics[width=.49\linewidth]{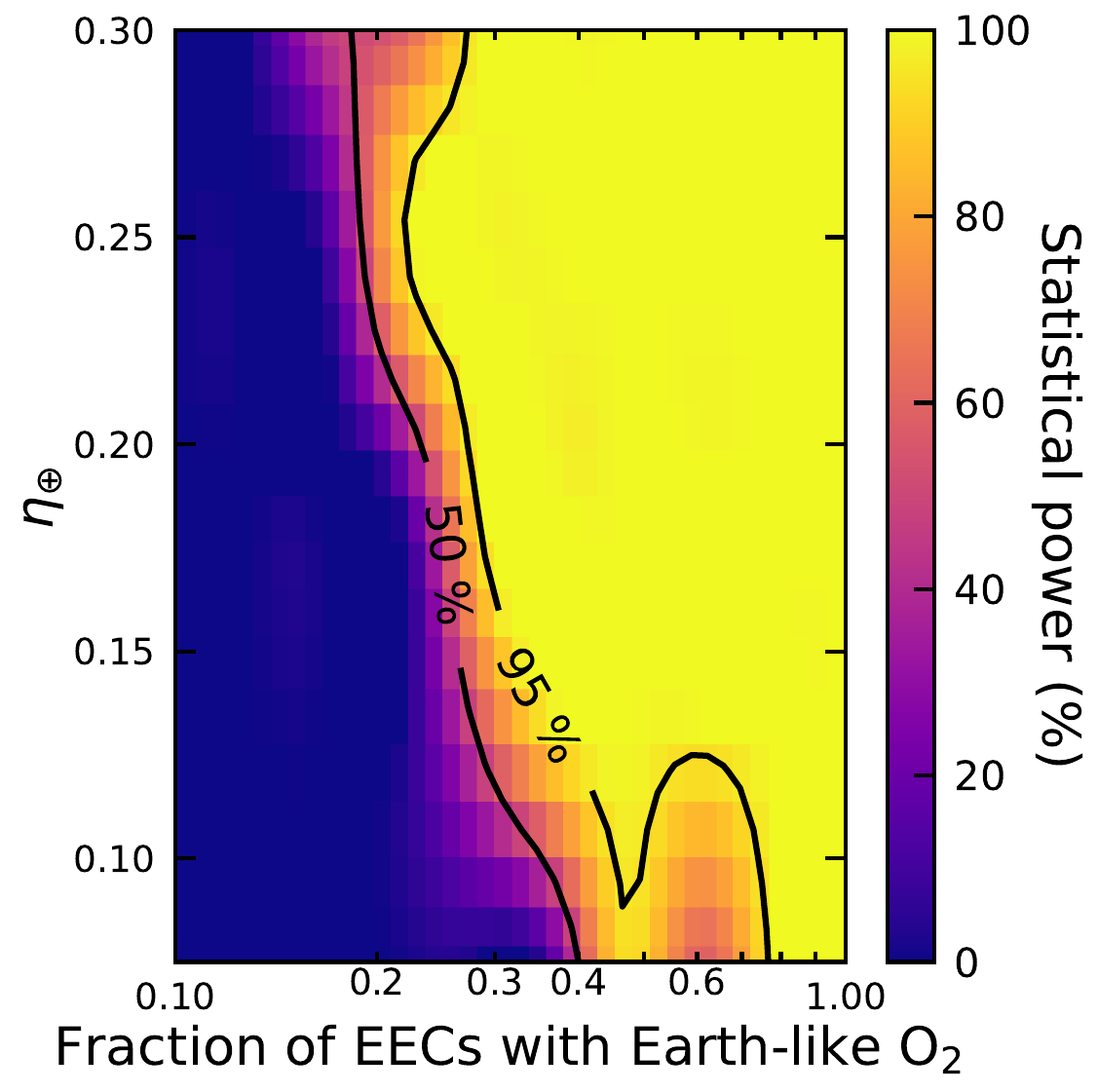}
\includegraphics[width=.49\linewidth]{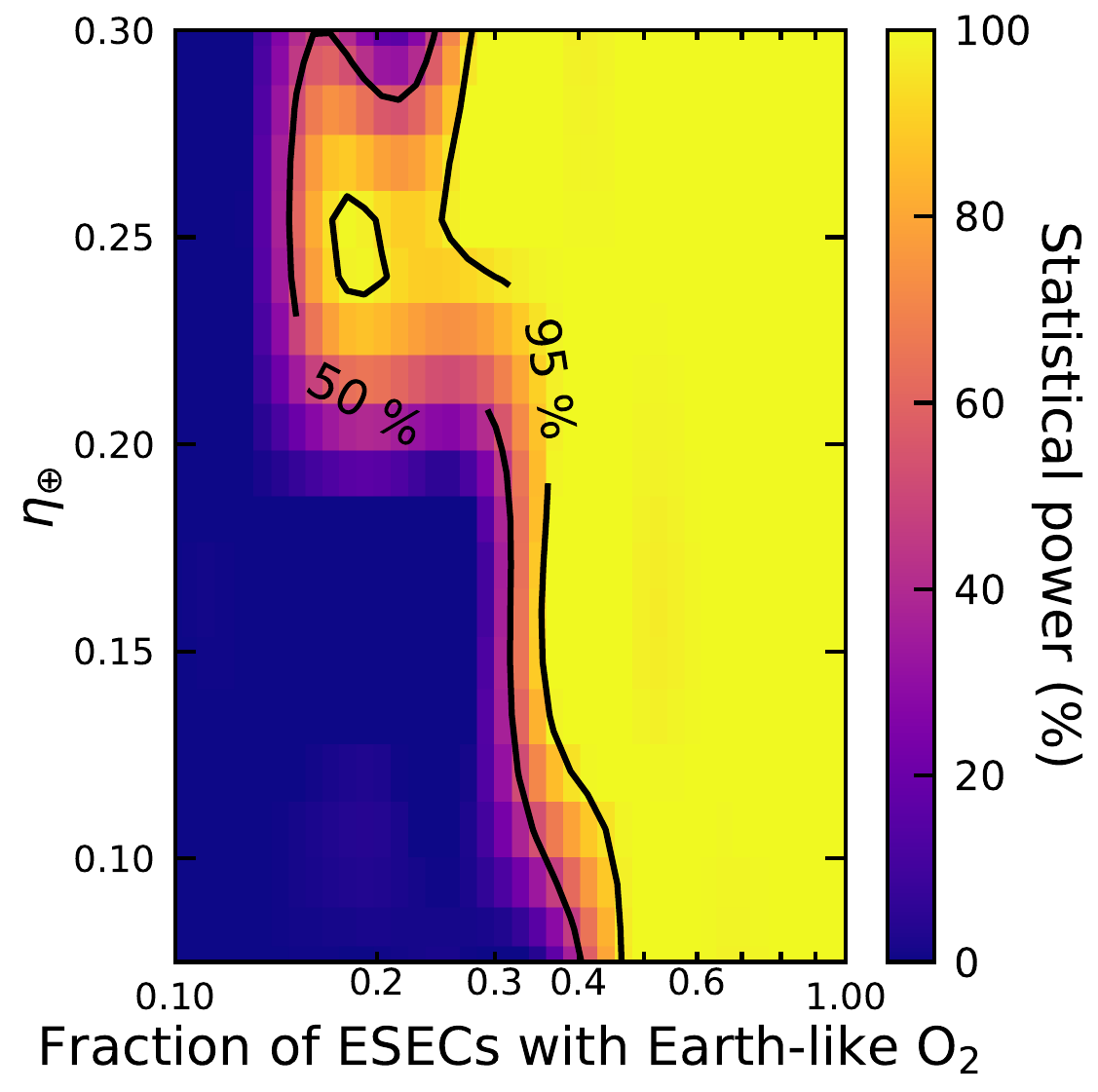}
\caption{Statistical power of the habitable zone oxygen hypothesis test on the ELT for different values of \ee\ and different fractions of exo-Earth candidates (EECs, left) and fractions of exo-super Earth candidates (ESECs, right) with Earth-like levels of \ot. Statistical power contours for 50\% and 95\% are shown in black. More planets with oxygen and/or higher values for \ee\ result in a greater ability to test this hypothesis, however, the inclusion of all super-Earths does not necessarily improve this ability. We note that these figures are super-sampled to highlight differences.\label{fig:o2-eta}}
\end{figure*}

In our test case, we aim to assess the statistical power to test the above hypothesis in a 10-year survey with the ELT considering different values of $f^{\mathrm{O}_\mathrm{2}}_{\mathrm{EEC}}$ (ten levels between 0.1 and 1) and \ee\ (five levels between 0.075 and 0.3). For simplicity, we set $f^{\mathrm{O}_\mathrm{2}}_{\text{non-EEC}} = 1\%$. A higher value of $f^{\mathrm{O}_\mathrm{2}}_{\text{non-EEC}}$ would inject more false-positive O$_2$ signals and skew our results to require fewer detections to test our hypothesis. We intentionally keep this value low because future surveys, like the one used in this hypothesis test, will likely only probe planets within the habitable zone defined either by incident stellar flux levels, or confirmed with liquid water detections on the habitable zone planets.
We imposed log-uniform priors $a_{\mathrm{inner}}=[0.1,2]$, $\Delta a=[0.01,3]$, $f_{\mathrm{HZ}}=[0.01,1]$, $f_{\text{non-HZ}}/f_{\mathrm{HZ}} = [0.01,1.0]$.

For each of these realizations, our planetary populations are still subject to intrinsic stochasticity that may impact the measured diagnostic power of the survey: planetary orbits and bulk properties are randomly drawn from distributions informed by exoplanet demographics, leading to random variations of the sample. To account for this while keeping computational costs reasonable, we ran ten randomized simulations per grid cell.

Figure~\ref{fig:o2-eta} shows the statistical power that can be achieved for different fractions of EECs with Earth-like levels of \ot\ and \ee. Our ability to test the habitable zone oxygen hypothesis is contingent on more than $\sim$1/3 of EECs having Earth-like levels of \ot\ if EECs are relatively common, and above $\sim$1/2 for low values of \ee. Figure~\ref{fig:sims-se} suggests that a 10-year survey of exo-super Earth candidates (ESECs) could probe about 1.5$\times$ more planets for Earth-like levels of \ot\ than a survey of only Earth-sized planets. As such, we ran the above test with the same parameters except considering ESECs rather than EECs. Figure~\ref{fig:o2-eta} shows that considering ESECs would not necessarily improve the ability to test the habitable zone oxygen hypothesis. One potential reason for this is that in a time-limited ground-based survey, a larger target sample would take longer to survey a significant fraction and test the hypothesis. A higher-resolution hypothesis grid with more samples per cell could test this more conclusively, but that is significantly more computationally expensive.

\begin{figure}[htbp!]
\centering
\includegraphics[width=\linewidth]{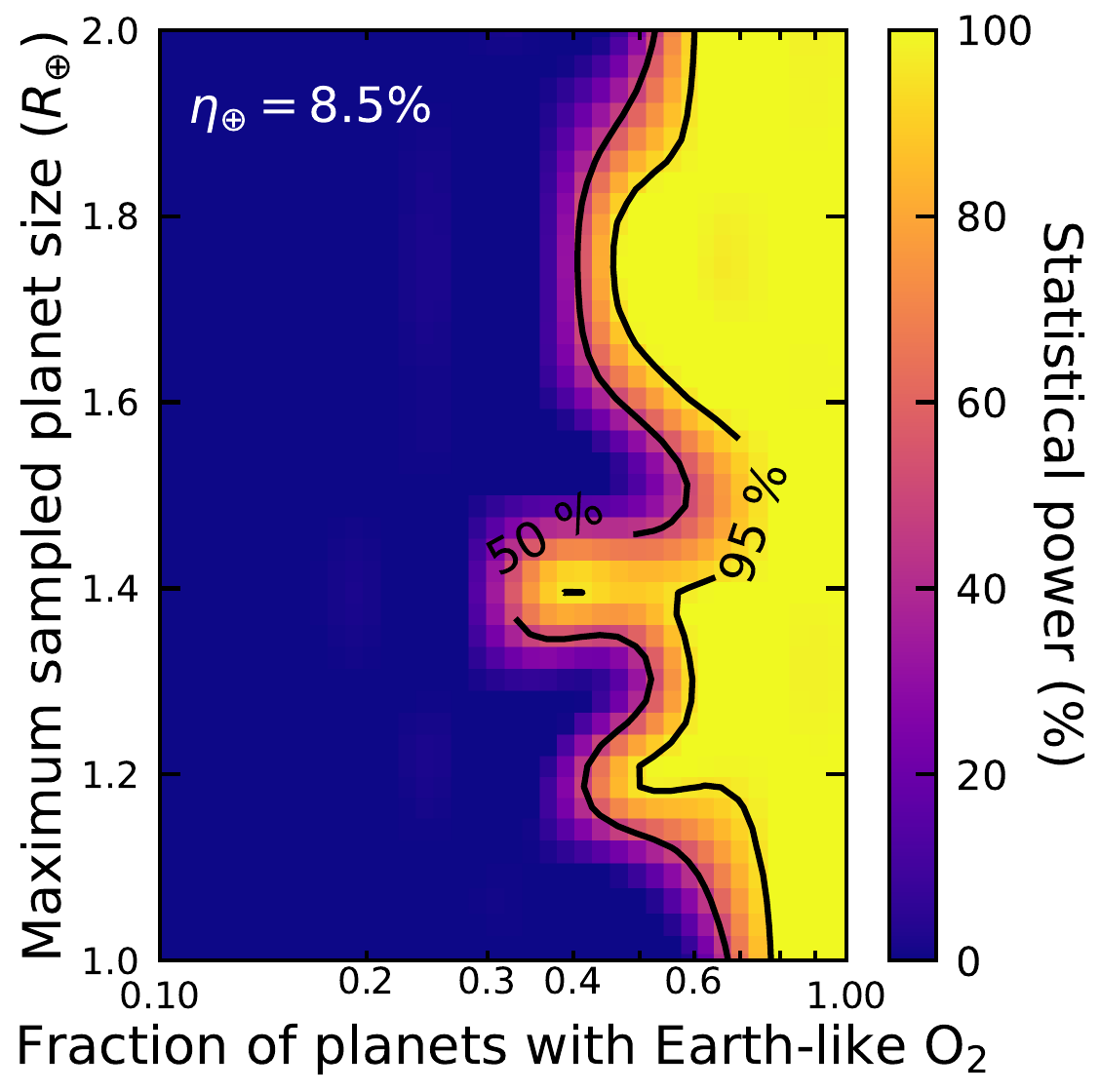}
\caption{Statistical power of the habitable zone oxygen hypothesis test on the ELT for different values of maximum sampled planet size and different fractions of planet candidates with Earth-like levels of \ot. Statistical power contours for 50\% and 95\% are shown in black. The maximum sampled planet size has minor effect on the statistical power between $\sim$1.2~$R_{\oplus}$ and $\sim$1.8~$R_{\oplus}$. We note that this figure is super-sampled. \label{fig:o2-rad}}
\end{figure}

In the previous two hypothesis tests, we considered $f^{\mathrm{O_2}}_{\mathrm{EEC}}$ or $f^{\mathrm{O_2}}_{\mathrm{ESEC}}$ for planets up to $1.4~R_{\oplus}$ (EEC) and $1.8~R_{\oplus}$ (ESEC) for different values of \ee. Next, instead of varying \ee, we set \ee\ = 8.5\% \citep[consistent with the most recent estimates of \ee\ for conservative habitable zone M dwarfs from][]{Bergsten2023} and tested the effect of maximum sampled planet size on our ability to test the habitable zone oxygen hypothesis. Figure~\ref{fig:o2-rad} shows the results of this test, indicating the maximum sampled planet size has only a small effect on the ability to test the habitable zone oxygen hypothesis between $\sim$1.2~$R_{\oplus}$ and 1.8~$R_{\oplus}$.

\section{Discussion}\label{sec:discussion}

The detection of \ot\ in the atmospheres of Earth-analogs, if found, will no doubt have profound impacts on astronomy and astrobiology. Our simulations show that Earth-like \ot\ levels could be probed on EECs with the GMT and ELT with planned instrumentation. There are already a half-dozen known probable super-Earths which may be ideal candidates to start searching for oxygen (Section~\ref{sec:known}), and planets like Proxima Centauri~b could be high-priority science targets for the GMT and ELT. Below we discuss simulation limitations and survey feasibility, \ot\, other biosignatures, and the importance of population-level biosignature interpretation, and end with a brief discussion of the TMT.

\subsection{Limitations and Survey Feasibility}\label{sec:limits}
Our simulations are based on loosely constrained (e.g., \ee) or unknown astrophysical parameters (e.g., orbital parameters). These contribute significantly to the uncertainties of our survey simulations. Exoplanet occurrence rate studies with TESS, Roman, and PLATO, and radial velocity surveys to find new EECs around nearby stars could help further reduce survey uncertainties and find new candidates for biosignature searches.

It is important to note that the GMT and ELT will be mostly sensitive to exoplanets orbiting M dwarf stars (Figure~\ref{fig:teff}), and particularly those in the habitable zone. These exoplanets will not be accessible to the Habitable Worlds Observatory, which is expected to primarily focus on Sun-like FGK stars \citep{Mamajek2024}. The nature of the habitability of M dwarf exoplanets is extensively studied in the literature \citep[e.g.,][]{Shields2013,Shields2016,Meadows2018a,Airapetian2020,Lobo2023} and is beyond the scope of this paper.

Since the extreme-AO systems for the ELTs are still in their planning stages, we rely on instrument parameters that are not yet finalized or that are based on current systems. Consequentially, our simulations can also be used to provide guidance to instrument teams as to the requirements for biosignature searches with the ELTs.

In Section~\ref{sec:survey}, we did not consider that multiple target planets could be observable at the same time, or by multiple ELTs. We also used the median calculated contrast ratio for each simulated planet (using only contrast ratios for times the planets were observable in one year of simulated observations). For simulation purposes, these are reasonable assumptions to provide upper limits to yields for an exoplanet biosignature survey for individual telescopes.

Our 10-year survey assumed we would observe during Q1 observing conditions. Since the ELTs are not solely dedicated to exoplanet science, it is unlikely that all Q1 observing time would be dedicated to the search for biosignatures. Though, exoplanets are not always observable or at ideal phases for direct imaging during Q1 observing conditions. As we assess in Appendix~\ref{sec:app1}, it makes little difference whether a survey is conducted in Q1 or Q2 observing conditions, which would likely be a factor in designing a biosignature survey. Until additional nearby EECs are identified, we can only report upper limits to survey yields for a hypothetical observing scenario as presented here.

\subsection{\texorpdfstring{\ot}{O2} as an Atmospheric Biosignature}\label{sec:o2}

Life on Earth has a complex evolutionary history with \ot. About 2.4 billion years ago, cyanobacteria took in solar energy and produced \ot\ as a waste product of oxygenic photosynthesis. Eventually, enough oxygen had accumulated in the ocean to begin escaping into the atmosphere during the ``Great Oxidation Event,'' ushering in the Proterozoic eon and the evolution of complex life. A second great atmospheric accumulation of oxygen occurred between about 800 and 600 million years ago, leading to present-day levels of oxygen and animal life \citep{Canfield2005,Kump2008,Sessions2009,Lyons2014,Lyons2021}. In this study, we focused on the ability to probe present-day Earth-like levels of \ot\ on rocky exoplanets. From 600 Mya to present day, Earth oxygen levels fluctuated between $\sim$10 and 30\%, settling at $\sim$21\% today \citep{Costa2014}. \citet{Chaloner1989} suggest that \ot\ concentrations of $\sim$35\% are the upper limit to sustain terrestrial vegetation due to the increased potential for combustibility. We leave the exploration of the limits of \ot\ detectability to future studies.

\subsection{Other Atmospheric Biosignatures}\label{sec:other-bio}

\ot\ on Earth is almost exclusively produced by oxygenic photosynthesis \citep{Segura2005}, and its present-day abundance, which is difficult to replicate by individual abiotic mechanisms \citep{Meadows2017}, makes it a favorable biosignature to search for in the atmospheres of exoplanets. However, there are still mechanisms that could produce an oxygen-rich atmosphere, such as water photolysis on a pure H$_{\mathrm{2}}$O world \citep{Wordsworth2014}. This necessitates robust elimination of false-positive abiotic \ot\ production scenarios. One way to do this is by observing multiple biosignatures \citep{Meadows2018}.

CH$_{\mathrm{4}}$ is another exoplanet biosignature candidate, as its primary production mechanism on Earth is from methanogenic bacteria in anaeorbic environments \citep{Segura2005,Thauer2008}. Abiotic CH$_{\mathrm{4}}$ can be produced by volcanic and geothermal processes, or by gas-water-rock reactions \citep{Etiope2013,Guzman-Marmolejo2013}, but on Earth, biological production of CH$_{\mathrm{4}}$ outweighs abiotic production by an order of magnitude \citep{Kelley2005,Segura2005}. The thermodynamic disequilibrium of CH$_{\mathrm{4}}$ and \ot\ on Earth \citep{Hitchcock1967,Sagan1993}, makes the disequilibrium pair a particularly enticing biosignature. Additionally, for M dwarf planets, which will be the primary targets for biosignature searches via direct imaging with the GMT and ELT, the photochemical lifetime of CH$_{\mathrm{4}}$ is $\sim$200 years, compared to $\sim$10 years on Earth \citep{Segura2005,Arney2019}, which might make it easier to detect on such planets.

CO$_{\mathrm{2}}$ and CH$_{\mathrm{4}}$ are another disequilibrium pair that may indicate life. This pair is particularly important for exoplanets with an Archean-like atmospheric composition when those two species dominated in the \ot\ deficient environment \citep{Haqq-Misra2008}. CH$_{\mathrm{4}}$ could be produced abiotically via mantle outgassing, but this would also produce abundant CO \citep{Krissansen-Totton2018}. A biologically active planet would likely have depleted CO levels because this would be the preferred energy source for methanogens over CO$_{\mathrm{2}}$, making the absence of CO a biological check on the CO$_{\mathrm{2}}$--CH$_{\mathrm{4}}$ pair \citep{Zahnle2011,Krissansen-Totton2018}.

\begin{figure*}[ht!]
\centering
\includegraphics[width=\linewidth]{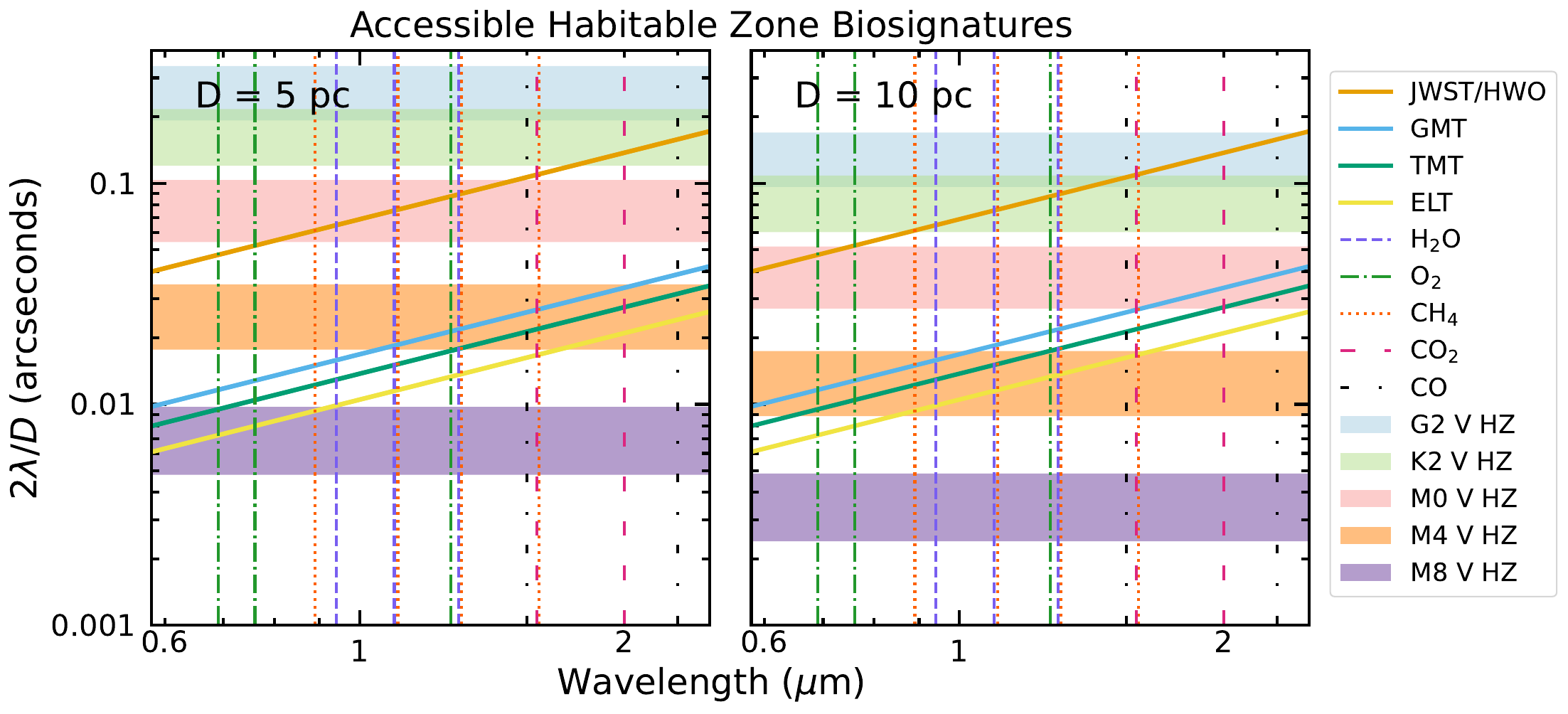}
\caption{The inner working angles at $2\lambda/D$ for 0.6--2~$\mu$m are plotted for a $\sim$6-meter JWST or anticipated HWO diameter aperture (orange solid line), the GMT (blue solid line), the TMT (green solid line), and the ELT (yellow solid line). The wavelengths of H$_{\mathrm{2}}$O (blue dashed), O$_{\mathrm{2}}$ (green dashed dotted), CH$_{\mathrm{4}}$ (orange dotted), CO$_{\mathrm{2}}$ (pink dashed), and CO (black dotted) for this wavelength region are plotted as vertical lines. At distances of 5~pc (left) and 10~pc (right), The shaded horizontal regions show the conservative habitable zones for representative G2~V through M8~V stars. If a habitable zone region falls above the inner working angle of a telescope, an exoplanet and its biosignatures should be observable at that wavelength range. For example, below 1~$\mu$m, the GMT should be able to observe O$_{\mathrm{2}}$, CH$_{\mathrm{4}}$, and H$_{\mathrm{2}}$O for planets in the habitable zones of a hypothetical M4~V star at 5~pc. The GMT will only be able to access biosignatures between 1 and 2~$\mu$m if planets fall in certain regions of the habitable zone of this hypothetical star, and it will not be able to access biosignatures beyond 2~$\mu$m. \label{fig:difflim}}
\end{figure*}

Figure~\ref{fig:difflim} shows the habitable zone accessibility of the biosignatures \ot, CH$_{\mathrm{4}}$, CO$_{\mathrm{2}}$, and CO at red-optical and near-infrared wavelengths for an inner-working angle of $2\lambda/D$ for the GMT, TMT, ELT, and JWST/HWO at distances of 5 and 10~pc. We have also included H$_{\mathrm{2}}$O, which is the basis of our definition of a habitable zone planet \citep{Kopparapu2014}. Our study focused on the \ot\ A-band at 0.76~$\mu$m, which could be probed in the atmospheres of most inner habitable zone planets orbiting stars up to M4~V by the GMT, TMT, and ELT out to $\sim$10~pc. Toward near-infrared wavelengths, the inner working angle decreases, diminishing the ability to probe some bands of biosignatures such as CO and CO$_{\mathrm{2}}$ in the atmospheres of exoplanets orbiting later-type and more distant stars. We leave the exploration of the ability to probe other biosignatures with the GMT, TMT, and ELT to future studies.

\subsection{A Statistical Interpretation of Biosignatures}\label{sec:statbio}

Even if it is possible to probe multiple biosignatures on an exoplanet, interpretation of detections/non-detections on a single planet will vary \citep[see e.g.,][]{Madhusudhan2023,Wogan2024,Tsai2024}. Therefore, it is crucial to survey a large number of planets to study biotic and abiotic processes and interpret biosignatures statistically \citep{Apai2019a}. Here, we focused on the potential to probe Earth-like levels of \ot\ on a statistical level on hypothetical nearby exoplanets using direct imaging and high-resolution spectroscopy. 
Studying exoplanets on a population level with large-aperture ground-based telescopes will allow testing of the habitable zone oxygen hypothesis and other statistical trends connected to small-planet atmosphere diversity and planet formation processes.

\subsection{Thirty Meter Telescope}\label{sec:tmt}

The TMT is the only 25--40-meter class telescope planned for the northern hemisphere. It should be noted that there are about 9,700 stars within 20~pc of the Sun brighter than 10th magnitude above a declination of 30$^{\circ}$. This is about 25\% of bright, nearby stars which are inaccessible to the GMT and ELT. A northern hemisphere large-aperture telescope is critical to maximize the search for biosignatures. Additionally, about 2/3 of the targets observable from a large-aperture telescope at 20$^{\circ}$ or 30$^{\circ}$ latitude overlap with the targets observable from the GMT and ELT in the southern hemisphere.

Having multiple facilities capable of searching for biosignatures is advantageous. When we are looking for S/N=5 biosignature detections over years of observations and hundreds of datasets, it will be important to have independent verification from, e.g., different telescopes and processing techniques to minimize false-positive scenarios. Multiple facilities can both expedite data collection on individual targets and potentially expand the target pool amenable to biosignature surveys if data from different instruments can be combined, and since weather conditions at one facility become less of an observing bottleneck. \citet{Hoeijmakers2018} successfully demonstrated combining more than 2,000 high-resolution reflected light spectra from different 2--8-meter class telescopes to measure the albedo of the hot Jupiter $\tau$ Bo\"{o}tis~b and achieved a planet-to-star contrast of $\sim$10$^{-5}$. The GMT, TMT, and ELT will be able to build upon these methods to reach smaller planets and potentially detect biosignatures.

The Planetary Systems Imager (PSI) is a planned second-generation instrument for the TMT which will have extreme-AO and high resolution spectroscopic capabilities \citep{Jensen-Clem2021,Jensen-Clem2022}. We did not model the potential of TMT+PSI in this paper because a detailed instrument model is not yet available. However, assuming similar performance to the ELT and GMT systems, the TMT could probe a $\sim$50\% larger sample than the GMT and -- equivalently -- probe a sample $\sim$66\% the size of the ELT sample in a 10 year survey, commensurate with the difference in telescope collecting area.
 
\section{Summary and Conclusions}\label{sec:summary}

The \texttt{Bioverse} survey simulation and hypothesis testing framework enables comprehensive and quantitative assessments of the capabilities of upcoming ground-based telescopes and space missions to study exoplanets. We expanded the \texttt{Bioverse} framework to assess the ability of the ELT and the GMT to probe Earth-like levels of \ot\ on nearby ($<$20~pc) habitable zone Earth-sized exoplanets via extreme-AO direct imaging and high resolution spectroscopy. The following list summarizes the main conclusions of this work:

\begin{itemize}
    
    \item Accounting for recent planet occurrence rate calculations, relative system velocities, and realistic target observability, we conducted a survey simulation to determine how many exo-Earths ($0.8S^{0.25} < R_p < 1.4~R_{\oplus}$) and super-Earths ($0.8S^{0.25} < R_p< 1.8~R_{\oplus}$) could be probed for Earth-like levels of \ot\ at S/N=5 with the GMT and ELT. In 10~years, the GMT could survey up to $\sim$7 and $\sim$10 Earths and super-Earths, respectively, while the ELT could survey up to $\sim$19 and $\sim$31 (Figures~\ref{fig:sims} and \ref{fig:sims-se}).
    
    \item Six known, habitable zone super-Earth candidates (Proxima Centauri~b, Ross~128~b, GJ~273~b, Wolf~1061~c, LTT~1445~A~d, and GJ~667~C~c) are ideal targets to search for \ot. Given favorable orbital and exoplanet atmosphere parameters, Earth-like levels of \ot\ could be probed in a matter of about one week to a few weeks of observing for four of these planets with the ELT, and about a month with the GMT (Figure~\ref{fig:sims-real}).

    \item A 10-year survey with the ELT could determine if habitable zone exo-Earths are more likely to have \ot. The diagnostic power of the survey to test this habitable zone oxygen hypothesis is sensitive to the fraction of EECs with Earth-like levels of \ot\ and the value of \ee, but not very sensitive to maximum planet size (Figures~\ref{fig:o2-eta} and \ref{fig:o2-rad}).

    \item The construction of a northern-hemisphere large-aperture telescope like the TMT would increase the potential target star sample by 25\%. Multiple facilities can provide independent verification of biosignature detections, expedite data collection on individual exoplanets, and potentially expand the exoplanet target sample on which to search for biosignatures.
\end{itemize}

The upcoming 25--40-meter class telescopes are going to be the first facilities capable of robustly detecting biosignatures with extreme-AO direct imaging and high resolution spectroscopy. This method is much more promising than transmission spectroscopy in the search for \ot\ \citep{Hardegree-Ullman2023}, and a handful of high-priority target planets are already available. The ability to validate biosignature detections and test the habitable zone oxygen hypothesis and other biosignature hypotheses, however, depends on probing a larger sample of planets than has been currently discovered.

\section{Acknowledgments}

We acknowledge the work of Bryan Wang to improve the \texttt{Bioverse} code. KKH-U acknowledges Megan Weiner Mansfield for general discussions about M~dwarf exoplanet atmospheres and Suri Rukdee for discussions about the LTT~1445~A system. We thank the anonymous referee for their helpful comments which significantly improved the clarity of this manuscript.

This material is based upon work supported by the National Aeronautics and Space Administration under Agreement No. 80NSSC21K0593 for the program ``Alien Earths.'' The results reported herein benefited from collaborations and/or information exchange within NASA's Nexus for Exoplanet System Science (NExSS) research coordination network sponsored by NASA’s Science Mission Directorate.

This material is based upon High Performance Computing (HPC) resources supported by the University of Arizona TRIF, UITS, and Research, Innovation, and Impact (RII) and maintained by the UArizona Research Technologies department.

This research has made use of NASA’s Astrophysics Data System.

\vspace{5mm}
\begin{large}\textit{Author contributions:}\end{large}
KKH-U and DA developed the project and planned its implementation. DA funded the project through the ``Alien Earths'' program and provided regular feedback and guidance as the project progressed. KKH-U updated the \texttt{Bioverse} code, performed the calculations and analysis, and drafted the manuscript. SYH created the GMT and ELT instrument models and computed contrast tables for \ot. MS assisted with \texttt{Bioverse} code issues and hypothesis test implementation. MK provided early models of PCS performance. MK and JK gave project input regarding the ELT and hypothesis testing. KW provided guidance on practical direct imaging observations. All authors contributed suggestions and edits to the manuscript.

\software{\texttt{astroplan} \citep{astroplan2018}, \texttt{astropy} \citep{astropy:2013, astropy:2018, AstropyCollaboration2022}, \texttt{Bioverse} \citep{Bixel2021,Hardegree-Ullman2023,Schlecker2024}, \texttt{dynesty} \citep{Speagle2020}, \texttt{matplotlib} \citep{Hunter:2007}, \texttt{numpy} \citep{harris2020array}, \texttt{pandas} \citep{mckinney-proc-scipy-2010,reback2020pandas}, \texttt{PyAstronomy} \citep{pya}}

\appendix
\section{Different Instrument Modes}\label{sec:app1}

For both the GMT and ELT, we first assessed different instrument setups and observing conditions. The GMagAO-X on GMT setup assumes a baseline number of 188 deformable mirror actuators, and we looked at two different AO control system optimization strategies: Max Speed, and Strehl Optimization. For PCS on the ELT, additional calculations were made for both a 128 or 200 actuator setup. Table~\ref{tab:GMTcontrast-full} shows results for the different AO control systems and observing conditions for the GMT, and Table~\ref{tab:ELTcontrast-full} shows results for different AO control systems, actuator counts, and observing conditions for the ELT. For each of these setups, we also give the end-to-end planet throughput for different host star magnitudes and angular separations for the GMT/GMagAO-X in Table~\ref{tab:GMTthroughput-full} and the ELT/PCS in Table~\ref{tab:ELTthroughput-full}.

\begin{figure*}[ht!]
\includegraphics[width=0.49\linewidth]{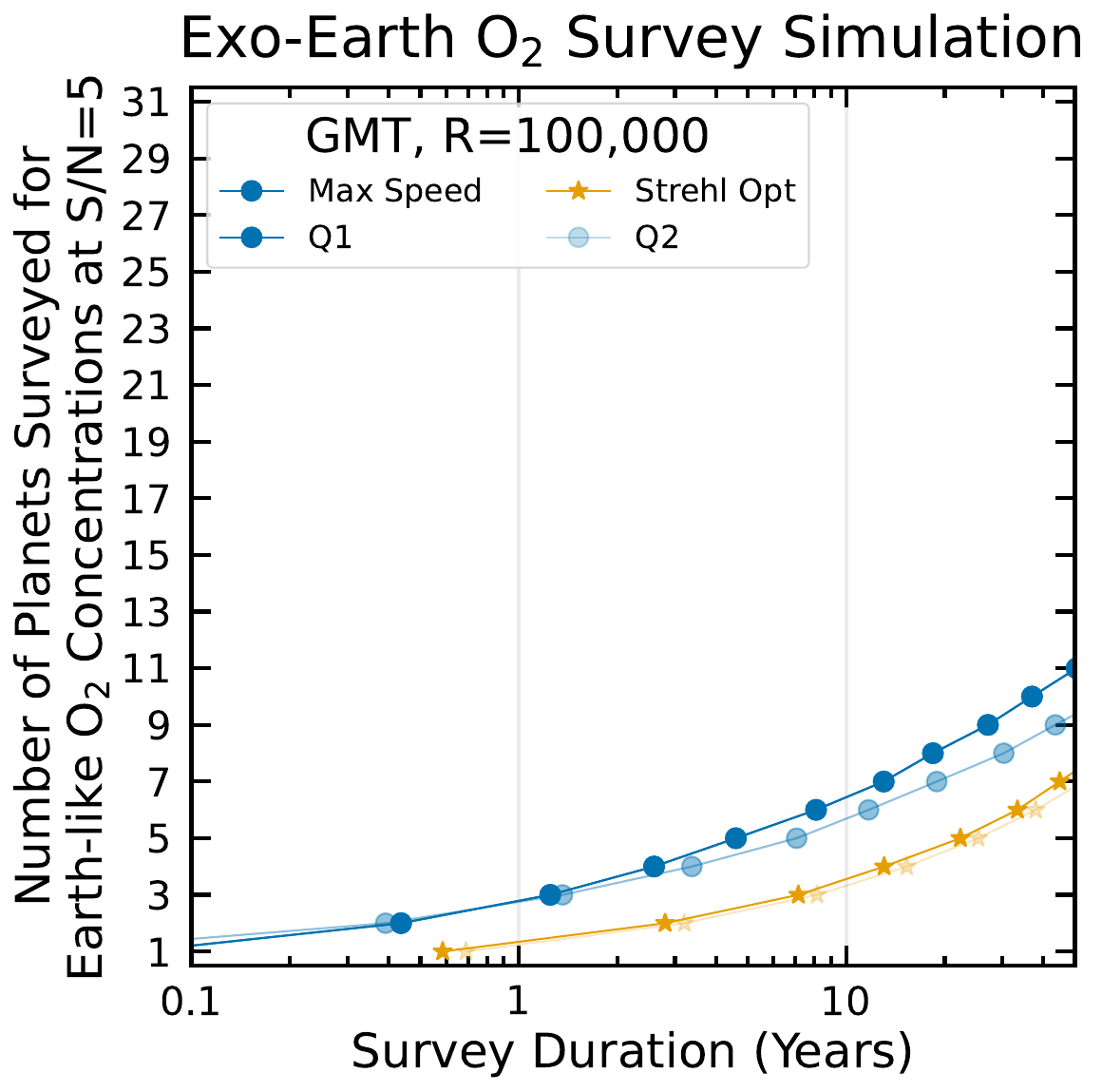}
\includegraphics[width=0.49\linewidth]{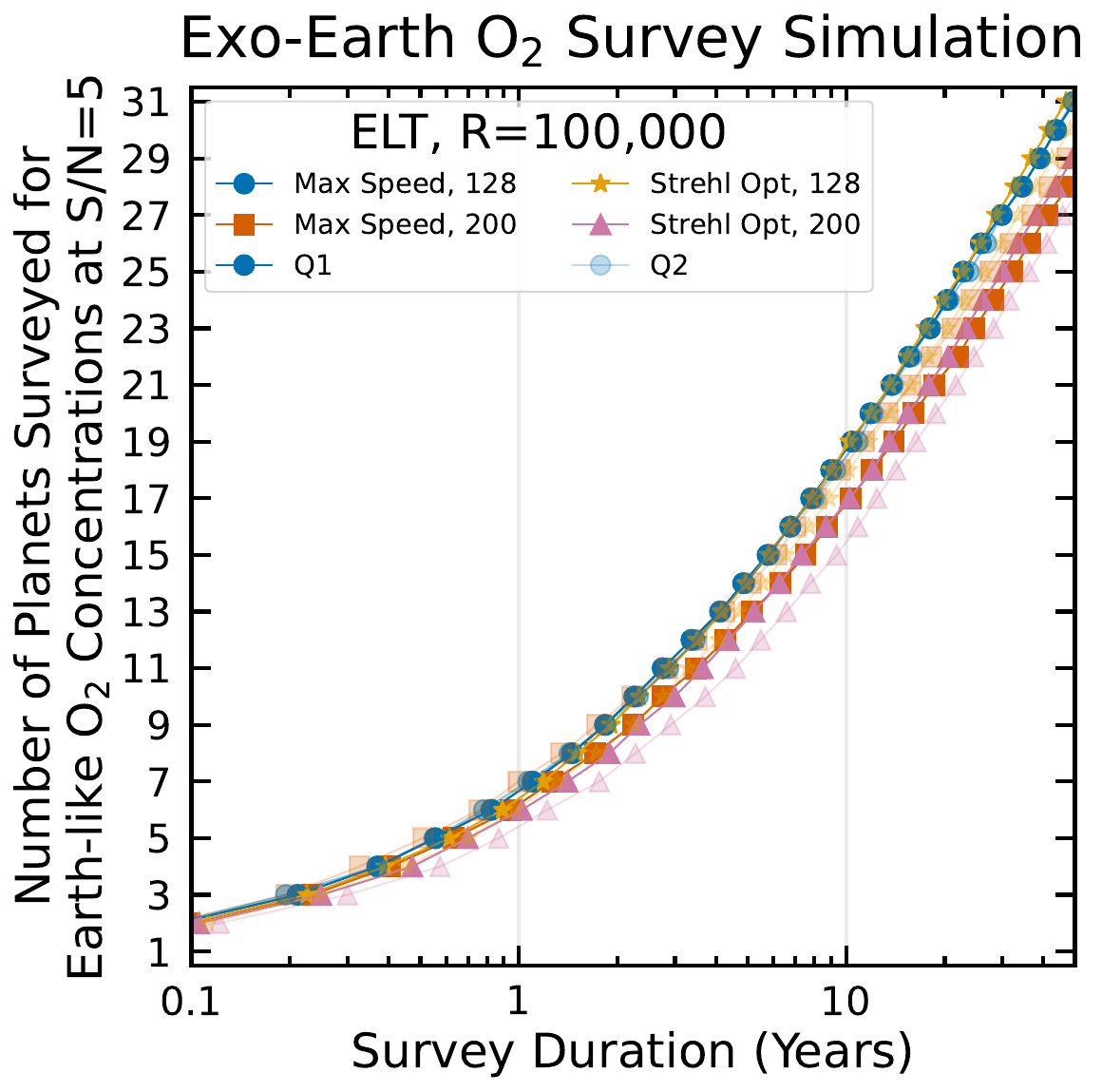} \\
\centering
\includegraphics[width=0.49\linewidth]{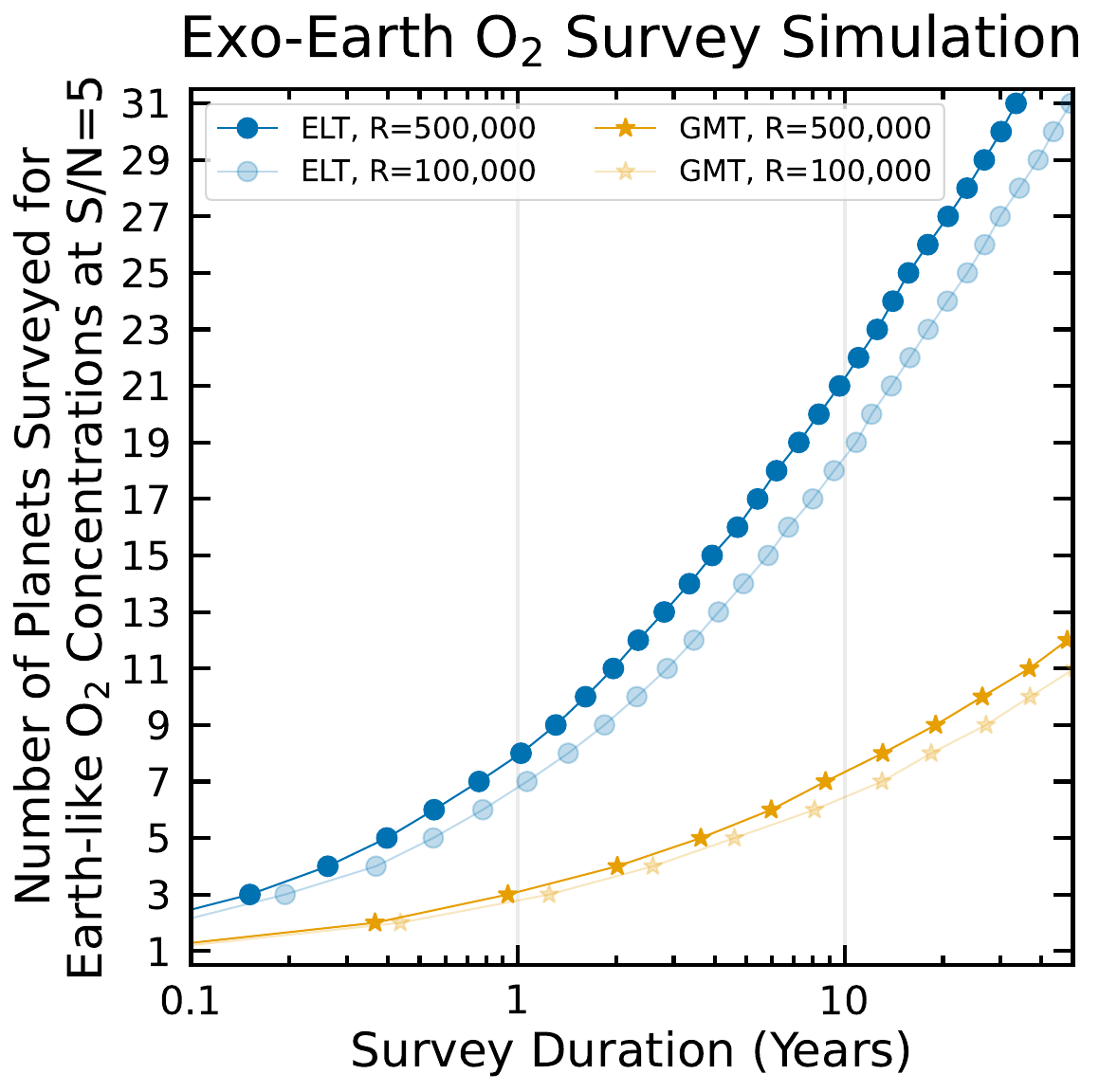} \\
\caption{Simulation results for a survey of Earth-like \ot\ levels at S/N=5 on exo-Earth candidates within 20~pc with different AO control systems and observing conditions for the GMT (upper left), additionally accounting for different deformable mirror actuator counts on the ELT (upper right), and considering an optimized setup for both the GMT and ELT but assessing the differences between R=100,000 and R=500,000 (lower panel). \label{fig:sims-app}}
\end{figure*}

Except for survey durations longer than 10 years on the ELT, Max Speed yields faster \ot\ detections than Strehl Optimization. There is a trade-off for observing in different conditions. Observing in Q2 conditions allows more observing time, but it will take longer to build up signal on a target. From these simulations, Q1 vs Q2 observing only makes marginal differences depending on the instrument setup. The large uncertainties (computed as the 16th and 84th percentile of the first through nth planets in the sorted universes; see Figure~\ref{fig:sims} for typical uncertainties for all models) effectively place all instrument setup results within 1$\sigma$. Nonetheless, we selected the GMT Max Speed Q1 and the ELT Max Speed, 128 actuators, Q1 instrument setups to perform additional optimistic simulations due to their marginally better performance than the other instrument setups for surveys with a duration of up to 10 years (Figure~\ref{fig:sims-app}).

\begin{table*}[!htbp]
\scriptsize
\begin{center}
\hskip-3.0cm\begin{tabular}{|c|cccccc|cccccc|} \hline
        & \multicolumn{6}{c|}{Max Speed, Q1}                    & \multicolumn{6}{c|}{Max Speed, Q2}                    \\ \cline{2-13}
Mag|$\theta$ & 5 mas & 10 mas & 15 mas & 30 mas & 50 mas & 120 mas & 5 mas & 10 mas & 15 mas & 30 mas & 50 mas & 120 mas \\ \hline
1 & 3.06E-08 & 6.09E-09 & 5.85E-09 & 4.53E-09 & 3.51E-09 & 2.35E-09 & 6.11E-08 & 1.27E-08 & 1.24E-08 & 9.52E-09 & 6.91E-09 & 4.04E-09 \\
3 & 9.79E-08 & 1.97E-08 & 1.93E-08 & 1.61E-08 & 1.32E-08 & 9.50E-09 & 1.68E-07 & 3.48E-08 & 3.42E-08 & 2.76E-08 & 2.17E-08 & 1.49E-08 \\
5 & 4.01E-07 & 8.10E-08 & 8.00E-08 & 6.88E-08 & 5.83E-08 & 4.39E-08 & 6.00E-07 & 1.25E-07 & 1.24E-07 & 1.07E-07 & 8.90E-08 & 6.54E-08 \\
7 & 2.26E-06 & 4.57E-07 & 4.54E-07 & 4.00E-07 & 3.47E-07 & 2.71E-07 & 3.26E-06 & 6.74E-07 & 6.76E-07 & 6.00E-07 & 5.19E-07 & 4.02E-07 \\
9 & 2.12E-05 & 4.30E-06 & 4.30E-06 & 3.87E-06 & 3.43E-06 & 2.72E-06 & 3.39E-05 & 6.99E-06 & 7.06E-06 & 6.44E-06 & 5.75E-06 & 4.62E-06 \\
10 & 8.87E-05 & 1.80E-05 & 1.81E-05 & 1.65E-05 & 1.49E-05 & 1.15E-05 & 1.71E-04 & 3.52E-05 & 3.57E-05 & 3.32E-05 & 3.03E-05 & 2.40E-05 \\ \hline
Mag     & \multicolumn{6}{c|}{Strehl Optimization, Q1}          & \multicolumn{6}{c|}{Strehl Optimization, Q2}          \\ \hline
1 & 3.88E-08 & 7.69E-09 & 7.33E-09 & 5.43E-09 & 3.92E-09 & 2.32E-09 & 6.11E-08 & 1.27E-08 & 1.24E-08 & 9.52E-09 & 6.91E-09 & 4.04E-09 \\
3 & 1.46E-07 & 2.90E-08 & 2.77E-08 & 2.05E-08 & 1.48E-08 & 8.67E-09 & 2.07E-07 & 4.32E-08 & 4.22E-08 & 3.25E-08 & 2.38E-08 & 1.45E-08 \\
5 & 7.34E-07 & 1.45E-07 & 1.38E-07 & 1.01E-07 & 7.15E-08 & 3.82E-08 & 1.04E-06 & 2.17E-07 & 2.11E-07 & 1.61E-07 & 1.15E-07 & 6.25E-08 \\
7 & 2.53E-06 & 5.01E-07 & 4.77E-07 & 3.52E-07 & 2.53E-07 & 1.46E-07 & 4.08E-06 & 8.51E-07 & 8.30E-07 & 6.33E-07 & 4.55E-07 & 2.56E-07 \\
9 & 1.34E-05 & 2.65E-06 & 2.52E-06 & 1.86E-06 & 1.32E-06 & 7.38E-07 & 1.83E-05 & 3.81E-06 & 3.72E-06 & 2.86E-06 & 2.09E-06 & 1.26E-06 \\
10 & 2.30E-05 & 4.55E-06 & 4.35E-06 & 3.29E-06 & 2.48E-06 & 1.65E-06 & 5.29E-05 & 1.10E-05 & 1.08E-05 & 8.23E-06 & 5.96E-06 & 3.35E-06 \\ \hline
\end{tabular}
\end{center}
\caption{GMT achievable $F_p/F_{\star}$ contrast at S/N=5 for 1 hour integration times using high contrast imaging with high-resolution spectroscopy for different host star magnitudes and planet-star angular separations ($\theta$). Tables are given for different AO control systems (Max Speed and Strehl Optimization) and Q1 and Q2 observing conditions.}\label{tab:GMTcontrast-full}
\end{table*}

\begin{table*}[!htbp]
\scriptsize
\begin{center}
\hskip-3.0cm\begin{tabular}{|c|cccccc|cccccc|} \hline
        & \multicolumn{6}{c|}{Max Speed, Q1}                    & \multicolumn{6}{c|}{Max Speed, Q2}                    \\ \cline{2-13}
Mag|$\theta$ & 5 mas & 10 mas & 15 mas & 30 mas & 50 mas & 120 mas & 5 mas & 10 mas & 15 mas & 30 mas & 50 mas & 120 mas \\ \hline
1 & 0.57 & 3.43 & 3.63 & 3.78 & 3.84 & 3.84 & 0.56 & 3.35 & 3.55 & 3.69 & 3.75 & 3.75 \\
3 & 0.57 & 3.40 & 3.60 & 3.74 & 3.80 & 3.81 & 0.55 & 3.30 & 3.50 & 3.64 & 3.69 & 3.70 \\
5 & 0.55 & 3.29 & 3.48 & 3.63 & 3.68 & 3.69 & 0.52 & 3.14 & 3.32 & 3.46 & 3.51 & 3.52 \\
7 & 0.48 & 2.90 & 3.07 & 3.19 & 3.24 & 3.25 & 0.43 & 2.59 & 2.74 & 2.85 & 2.89 & 2.90 \\
9 & 0.30 & 1.78 & 1.88 & 1.96 & 1.99 & 1.99 & 0.21 & 1.24 & 1.31 & 1.36 & 1.38 & 1.39 \\
10 & 0.16 & 0.94 & 1.00 & 1.04 & 1.06 & 1.06 & 0.08 & 0.48 & 0.51 & 0.53 & 0.53 & 0.54 \\ \hline
Mag     & \multicolumn{6}{c|}{Strehl Optimization, Q1}          & \multicolumn{6}{c|}{Strehl Optimization, Q2}          \\ \hline
1 & 0.57 & 3.43 & 3.63 & 3.78 & 3.84 & 3.84 & 0.56 & 3.35 & 3.55 & 3.69 & 3.75 & 3.75 \\
3 & 0.57 & 3.41 & 3.61 & 3.76 & 3.82 & 3.82 & 0.55 & 3.31 & 3.50 & 3.65 & 3.70 & 3.71 \\
5 & 0.56 & 3.36 & 3.55 & 3.70 & 3.76 & 3.76 & 0.54 & 3.22 & 3.41 & 3.55 & 3.60 & 3.61 \\
7 & 0.54 & 3.23 & 3.42 & 3.56 & 3.62 & 3.62 & 0.50 & 2.99 & 3.17 & 3.29 & 3.35 & 3.35 \\
9 & 0.49 & 2.92 & 3.09 & 3.22 & 3.27 & 3.27 & 0.41 & 2.47 & 2.62 & 2.73 & 2.77 & 2.77 \\
10 & 0.44 & 2.64 & 2.80 & 2.91 & 2.96 & 2.96 & 0.34 & 2.06 & 2.18 & 2.26 & 2.30 & 2.30 \\ \hline
\end{tabular}
\end{center}
\caption{End-to-end GMagAO-X planet throughput in percentages (from 0\% to 100\%) for different host star magnitudes and planet-star angular separations ($\theta$). Tables are given for different AO control systems (Max Speed and Strehl Optimization) and Q1 and Q2 observing conditions.}\label{tab:GMTthroughput-full}
\end{table*}

\begin{table*}[htb!]
\begin{center}
\scriptsize
\begin{tabular}{|c|cccccc|cccccc|} \hline
 & \multicolumn{6}{c|}{Max Speed, 128 actuators, Q1}                & \multicolumn{6}{c|}{Max Speed, 128 actuators, Q2}                \\ \cline{2-13}
Mag|$\theta$ & 5 mas    & 10 mas   & 15 mas   & 30 mas   & 50 mas   & 120 mas  & 5 mas    & 10 mas   & 15 mas   & 30 mas   & 50 mas   & 120 mas  \\ \hline
1          & 2.55E-09 & 2.08E-09 & 1.99E-09 & 1.55E-09 & 1.25E-09 & 9.36E-10 & 5.00E-09 & 4.07E-09 & 3.89E-09 & 3.11E-09 & 2.58E-09 & 1.97E-09 \\
3          & 7.12E-09 & 5.84E-09 & 5.64E-09 & 4.64E-09 & 3.90E-09 & 2.95E-09 & 1.30E-08 & 1.06E-08 & 1.02E-08 & 8.33E-09 & 7.06E-09 & 5.50E-09 \\
5          & 2.48E-08 & 2.04E-08 & 1.99E-08 & 1.68E-08 & 1.43E-08 & 1.09E-08 & 3.91E-08 & 3.21E-08 & 3.12E-08 & 2.64E-08 & 2.26E-08 & 1.75E-08 \\
7          & 1.03E-07 & 8.49E-08 & 8.30E-08 & 7.14E-08 & 6.17E-08 & 4.78E-08 & 1.49E-07 & 1.22E-07 & 1.19E-07 & 1.02E-07 & 8.82E-08 & 6.85E-08 \\
9          & 5.72E-07 & 4.72E-07 & 4.64E-07 & 4.06E-07 & 3.57E-07 & 2.82E-07 & 7.95E-07 & 6.54E-07 & 6.41E-07 & 5.60E-07 & 4.91E-07 & 3.90E-07 \\
10         & 1.62E-06 & 1.34E-06 & 1.32E-06 & 1.17E-06 & 1.03E-06 & 8.24E-07 & 2.34E-06 & 1.93E-06 & 1.89E-06 & 1.67E-06 & 1.49E-06 & 1.20E-06 \\ \hline
Mag  & \multicolumn{6}{c|}{Max Speed, 200 actuators, Q1}                & \multicolumn{6}{c|}{Max Speed, 200 actuators, Q2}                \\ \hline
1          & 2.24E-09 & 1.83E-09 & 1.72E-09 & 1.27E-09 & 9.48E-10 & 6.18E-10 & 4.06E-09 & 3.33E-09 & 3.13E-09 & 2.27E-09 & 1.66E-09 & 1.01E-09 \\
3          & 6.39E-09 & 5.24E-09 & 5.03E-09 & 4.02E-09 & 3.27E-09 & 2.31E-09 & 1.07E-08 & 8.77E-09 & 8.32E-09 & 6.31E-09 & 4.92E-09 & 3.39E-09 \\
5          & 2.34E-08 & 1.93E-08 & 1.87E-08 & 1.56E-08 & 1.31E-08 & 9.73E-09 & 3.39E-08 & 2.79E-08 & 2.70E-08 & 2.21E-08 & 1.83E-08 & 1.33E-08 \\
7          & 1.07E-07 & 8.78E-08 & 8.58E-08 & 7.37E-08 & 6.35E-08 & 4.89E-08 & 1.46E-07 & 1.20E-07 & 1.17E-07 & 9.95E-08 & 8.50E-08 & 6.46E-08 \\
9          & 7.13E-07 & 5.88E-07 & 5.77E-07 & 5.07E-07 & 4.46E-07 & 3.52E-07 & 9.95E-07 & 8.22E-07 & 8.07E-07 & 7.07E-07 & 6.21E-07 & 4.93E-07 \\
10         & 2.22E-06 & 1.83E-06 & 1.80E-06 & 1.60E-06 & 1.42E-06 & 1.13E-06 & 3.42E-06 & 2.82E-06 & 2.78E-06 & 2.48E-06 & 2.21E-06 & 1.78E-06 \\ \hline
Mag  & \multicolumn{6}{c|}{Strehl Optimization, 128 actuators, Q1}      & \multicolumn{6}{c|}{Strehl Optimization, 128 actuators, Q2}      \\ \hline
1          & 2.55E-09 & 2.08E-09 & 1.99E-09 & 1.55E-09 & 1.25E-09 & 9.36E-10 & 5.00E-09 & 4.07E-09 & 3.89E-09 & 3.11E-09 & 2.58E-09 & 1.97E-09 \\
3          & 8.29E-09 & 6.78E-09 & 6.44E-09 & 4.94E-09 & 3.91E-09 & 2.86E-09 & 1.57E-08 & 1.28E-08 & 1.21E-08 & 9.37E-09 & 7.48E-09 & 5.46E-09 \\
5          & 3.03E-08 & 2.48E-08 & 2.35E-08 & 1.76E-08 & 1.36E-08 & 9.60E-09 & 5.62E-08 & 4.57E-08 & 4.30E-08 & 3.20E-08 & 2.44E-08 & 1.65E-08 \\
7          & 1.46E-07 & 1.20E-07 & 1.13E-07 & 8.14E-08 & 5.90E-08 & 3.59E-08 & 1.88E-07 & 1.53E-07 & 1.44E-07 & 1.07E-07 & 8.11E-08 & 5.57E-08 \\
9          & 5.94E-07 & 4.85E-07 & 4.57E-07 & 3.30E-07 & 2.40E-07 & 1.49E-07 & 9.28E-07 & 7.53E-07 & 7.06E-07 & 5.08E-07 & 3.68E-07 & 2.24E-07 \\
10         & 1.20E-06 & 9.77E-07 & 9.20E-07 & 6.70E-07 & 4.93E-07 & 3.18E-07 & 1.81E-06 & 1.47E-06 & 1.38E-06 & 9.99E-07 & 7.34E-07 & 4.70E-07 \\ \hline
Mag  & \multicolumn{6}{c|}{Strehl Optimization, 200 actuators, Q1}      & \multicolumn{6}{c|}{Strehl Optimization, 200 actuators, Q2}      \\ \hline
1          & 2.91E-09 & 2.37E-09 & 2.23E-09 & 1.59E-09 & 1.13E-09 & 6.42E-10 & 4.06E-09 & 3.33E-09 & 3.13E-09 & 2.27E-09 & 1.66E-09 & 1.01E-09 \\
3          & 1.09E-08 & 8.92E-09 & 8.37E-09 & 5.96E-09 & 4.21E-09 & 2.33E-09 & 1.36E-08 & 1.11E-08 & 1.05E-08 & 7.58E-09 & 5.50E-09 & 3.35E-09 \\
5          & 3.68E-08 & 3.01E-08 & 2.82E-08 & 2.02E-08 & 1.44E-08 & 8.38E-09 & 5.10E-08 & 4.18E-08 & 3.93E-08 & 2.83E-08 & 2.04E-08 & 1.21E-08 \\
7          & 1.40E-07 & 1.14E-07 & 1.07E-07 & 7.72E-08 & 5.56E-08 & 3.33E-08 & 2.56E-07 & 2.10E-07 & 1.97E-07 & 1.40E-07 & 9.86E-08 & 5.33E-08 \\
9          & 7.10E-07 & 5.79E-07 & 5.44E-07 & 3.88E-07 & 2.76E-07 & 1.56E-07 & 1.06E-06 & 8.70E-07 & 8.18E-07 & 5.84E-07 & 4.13E-07 & 2.29E-07 \\
10         & 1.53E-06 & 1.25E-06 & 1.17E-06 & 8.40E-07 & 5.99E-07 & 3.46E-07 & 2.19E-06 & 1.79E-06 & 1.69E-06 & 1.21E-06 & 8.64E-07 & 4.99E-07 \\ \hline
\end{tabular}
\caption{ELT achievable $F_p/F_{\star}$ contrast at S/N=5 for 1 hour integration times using high contrast imaging with high-resolution spectroscopy for different host star magnitudes and planet-star angular separations ($\theta$). Tables are given for different AO control systems (Max Speed and Strehl Optimization), 128 and 200 deformable mirror actuators, and Q1 and Q2 observing conditions.}\label{tab:ELTcontrast-full}
\end{center}
\end{table*}

\begin{table*}[htb!]
\begin{center}
\scriptsize
\begin{tabular}{|c|cccccc|cccccc|} \hline
 & \multicolumn{6}{c|}{Max Speed, 128 actuators, Q1}                & \multicolumn{6}{c|}{Max Speed, 128 actuators, Q2}                \\ \cline{2-13}
Mag|$\theta$ & 5 mas    & 10 mas   & 15 mas   & 30 mas   & 50 mas   & 120 mas  & 5 mas    & 10 mas   & 15 mas   & 30 mas   & 50 mas   & 120 mas  \\ \hline
 1 & 2.74 & 3.88 & 3.92 & 3.95 & 3.96 & 3.96 & 2.32 & 3.29 & 3.33 & 3.36 & 3.36 & 3.36 \\
3 & 2.73 & 3.87 & 3.92 & 3.95 & 3.95 & 3.95 & 2.32 & 3.28 & 3.32 & 3.35 & 3.35 & 3.35 \\
5 & 2.72 & 3.86 & 3.90 & 3.94 & 3.94 & 3.94 & 2.30 & 3.26 & 3.30 & 3.33 & 3.33 & 3.34 \\
7 & 2.69 & 3.80 & 3.85 & 3.88 & 3.88 & 3.89 & 2.26 & 3.20 & 3.24 & 3.26 & 3.27 & 3.27 \\
9 & 2.48 & 3.51 & 3.55 & 3.59 & 3.59 & 3.59 & 2.00 & 2.84 & 2.87 & 2.90 & 2.90 & 2.90 \\
10 & 2.13 & 3.01 & 3.05 & 3.07 & 3.08 & 3.08 & 1.57 & 2.22 & 2.25 & 2.27 & 2.27 & 2.27 \\ \hline
Mag  & \multicolumn{6}{c|}{Max Speed, 200 actuators, Q1}                & \multicolumn{6}{c|}{Max Speed, 200 actuators, Q2}                \\ \hline
1 & 2.94 & 4.16 & 4.21 & 4.25 & 4.25 & 4.25 & 2.74 & 3.88 & 3.93 & 3.96 & 3.97 & 3.97 \\
3 & 2.93 & 4.15 & 4.20 & 4.24 & 4.24 & 4.24 & 2.73 & 3.87 & 3.92 & 3.95 & 3.95 & 3.95 \\
5 & 2.91 & 4.12 & 4.17 & 4.21 & 4.21 & 4.21 & 2.70 & 3.83 & 3.88 & 3.91 & 3.91 & 3.91 \\
7 & 2.82 & 4.00 & 4.04 & 4.08 & 4.08 & 4.08 & 2.58 & 3.65 & 3.70 & 3.73 & 3.73 & 3.73 \\
9 & 2.37 & 3.36 & 3.40 & 3.43 & 3.43 & 3.43 & 1.92 & 2.73 & 2.76 & 2.78 & 2.79 & 2.79 \\
10 & 1.88 & 2.66 & 2.69 & 2.71 & 2.72 & 2.72 & 1.26 & 1.78 & 1.80 & 1.82 & 1.82 & 1.82 \\
\hline
Mag  & \multicolumn{6}{c|}{Strehl Optimization, 128 actuators, Q1}      & \multicolumn{6}{c|}{Strehl Optimization, 128 actuators, Q2}      \\ \hline
1 & 2.74 & 3.88 & 3.92 & 3.95 & 3.96 & 3.96 & 2.32 & 3.29 & 3.33 & 3.36 & 3.36 & 3.36 \\
3 & 2.73 & 3.87 & 3.92 & 3.95 & 3.95 & 3.95 & 2.32 & 3.28 & 3.32 & 3.35 & 3.35 & 3.35 \\
5 & 2.73 & 3.86 & 3.91 & 3.94 & 3.94 & 3.95 & 2.31 & 3.27 & 3.31 & 3.34 & 3.34 & 3.34 \\
7 & 2.71 & 3.84 & 3.89 & 3.92 & 3.92 & 3.92 & 2.28 & 3.24 & 3.27 & 3.30 & 3.30 & 3.31 \\
9 & 2.67 & 3.78 & 3.82 & 3.86 & 3.86 & 3.86 & 2.22 & 3.14 & 3.18 & 3.20 & 3.21 & 3.21 \\
10 & 2.62 & 3.71 & 3.76 & 3.79 & 3.79 & 3.79 & 2.14 & 3.04 & 3.07 & 3.10 & 3.10 & 3.10 \\
\hline
Mag  & \multicolumn{6}{c|}{Strehl Optimization, 200 actuators, Q1}      & \multicolumn{6}{c|}{Strehl Optimization, 200 actuators, Q2}      \\ \hline
1 & 2.94 & 4.16 & 4.21 & 4.25 & 4.25 & 4.25 & 2.74 & 3.88 & 3.93 & 3.96 & 3.97 & 3.97 \\
3 & 2.93 & 4.15 & 4.20 & 4.24 & 4.24 & 4.24 & 2.73 & 3.87 & 3.92 & 3.95 & 3.96 & 3.96 \\
5 & 2.92 & 4.14 & 4.19 & 4.22 & 4.23 & 4.23 & 2.71 & 3.85 & 3.89 & 3.92 & 3.93 & 3.93 \\
7 & 2.89 & 4.09 & 4.14 & 4.18 & 4.18 & 4.18 & 2.66 & 3.77 & 3.81 & 3.85 & 3.85 & 3.85 \\
9 & 2.80 & 3.97 & 4.02 & 4.06 & 4.06 & 4.06 & 2.51 & 3.56 & 3.60 & 3.64 & 3.64 & 3.64 \\
10 & 2.71 & 3.84 & 3.88 & 3.92 & 3.92 & 3.92 & 2.35 & 3.33 & 3.37 & 3.40 & 3.40 & 3.40 \\
\hline
\end{tabular}
\caption{End-to-end ELT/PCS planet throughput in percentages (from 0\% to 100\%) for different host star magnitudes and planet-star angular separations ($\theta$). Tables are given for different AO control systems (Max Speed and Strehl Optimization), 128 and 200 deformable mirror actuators, and Q1 and Q2 observing conditions.}\label{tab:ELTthroughput-full}
\end{center}
\end{table*}

A higher resolution spectrograph produces narrower and deeper absorption and emission lines. For \ot, higher resolution gives access to more relative system velocity space. We modeled an $R=500,000$ spectrograph coupled to our AO systems, for which severe line blending in the \ot\ A-band occurs when $|\Delta \mathrm{RV}|<10$ km s$^{-1}$ and $40\ \mathrm{km\ s}^{-1} < |\Delta \mathrm{RV}| < 48\ \mathrm{km\ s}^{-1}$ \citep{Lopez-Morales2019,Hardegree-Ullman2023}. As noted by \citet{Fowler2023}, post-processing efficiency does not change at $R>100,000$, so we are able to use Tables~\ref{tab:GMTcontrast-full} and \ref{tab:ELTcontrast-full} for our $R=500,000$ calculations. The lower panel of Figure~\ref{fig:sims-app} shows the results for $R=500,000$ compared to $R=100,000$, with a $\sim$1 or 2 planet increase in planets surveyed for a survey of 10 years, suggesting a higher resolution spectrograph is not necessarily a good investment for this science case.

\clearpage
\bibliography{main}{}
\bibliographystyle{aasjournal}

\end{document}